%% file: main.tex
\documentclass[acmsmall]{acmart}  %

\AtBeginDocument{%
  }

\setcopyright{acmlicensed}
\copyrightyear{2026}
\acmYear{2026}
\acmDOI{XXXXXXX.XXXXXXX}

\acmJournal{JACM}
\acmVolume{37}
\acmNumber{4}
\acmArticle{111}
\acmMonth{8}

\usepackage{enumitem}            %
\usepackage{tabu}                %

\begin{document}

\title{Calibrating Trustworthiness: Co-Designing Metrics and Visualizations for Evaluating LLMs in Education}

\author{Adam Coscia}
\orcid{0000-0002-0429-9295}
\email{acoscia6@gatech.edu}
\affiliation{%
  \institution{Georgia Institute of Technology}
  \city{Atlanta}
  \state{Georgia}
  \country{USA}
}%

\author{Sujata Duwal}
\orcid{0009-0004-5764-690X}
\email{sduwal3@gatech.edu}
\affiliation{%
  \institution{Georgia Institute of Technology}
  \city{Atlanta}
  \state{Georgia}
  \country{USA}
}%

\author{Langdon Holmes}
\orcid{0000-0003-4338-4609}
\email{langdon.holmes@vanderbilt.edu}
\affiliation{%
  \institution{Vanderbilt University}
  \streetaddress{2201 West End Ave}
  \city{Nashville}
  \state{Tennessee}
  \country{USA}
  \postcode{37235}
}%

\author{Scott Crossley}
\orcid{0000-0002-5148-0273}
\email{scott.crossley@vanderbilt.edu}
\affiliation{%
  \institution{Vanderbilt University}
  \streetaddress{2201 West End Ave}
  \city{Nashville}
  \state{Tennessee}
  \country{USA}
  \postcode{37235}
}%

\author{Alex Endert}
\orcid{0000-0002-6914-610X}
\email{endert@gatech.edu}
\affiliation{%
  \institution{Georgia Institute of Technology}
  \city{Atlanta}
  \state{Georgia}
  \country{USA}
}%

\renewcommand{\shortauthors}{Coscia et al.}

\begin{abstract}
  LLMs are reshaping educational technology, yet evaluating their responses for pedagogical alignment remains underexplored, relying heavily on the expertise of learning engineers building the technology.
  To bridge this gap, we explore trustworthiness as a structured lens for evaluation, leveraging existing measures of LLM trustworthiness to systematically identify potential pedagogical disruptions.
  Through a longitudinal co-design process with learning engineers developing an LLM-powered digital textbook, we: (1) co-constructed five trustworthiness metrics comprising 20 measures tailored to pedagogical use; (2) designed visualizations that map trustworthiness violations onto LLM responses; and (3) evaluated how these tools help learning engineers make A/B comparisons of LLM responses.
  Making trustworthiness explicit increased inter-rater reliability while helping learning engineers resolve conflicting objectives and produce more consistent judgments.
  We discuss the emergent benefits of trustworthiness as a lens for evaluating LLMs in education and propose new design guidelines for future evaluation tools that enable pedagogically-aligned, LLM-powered learning tools.
\end{abstract}



\begin{teaserfigure}
  \includegraphics[width=\linewidth]{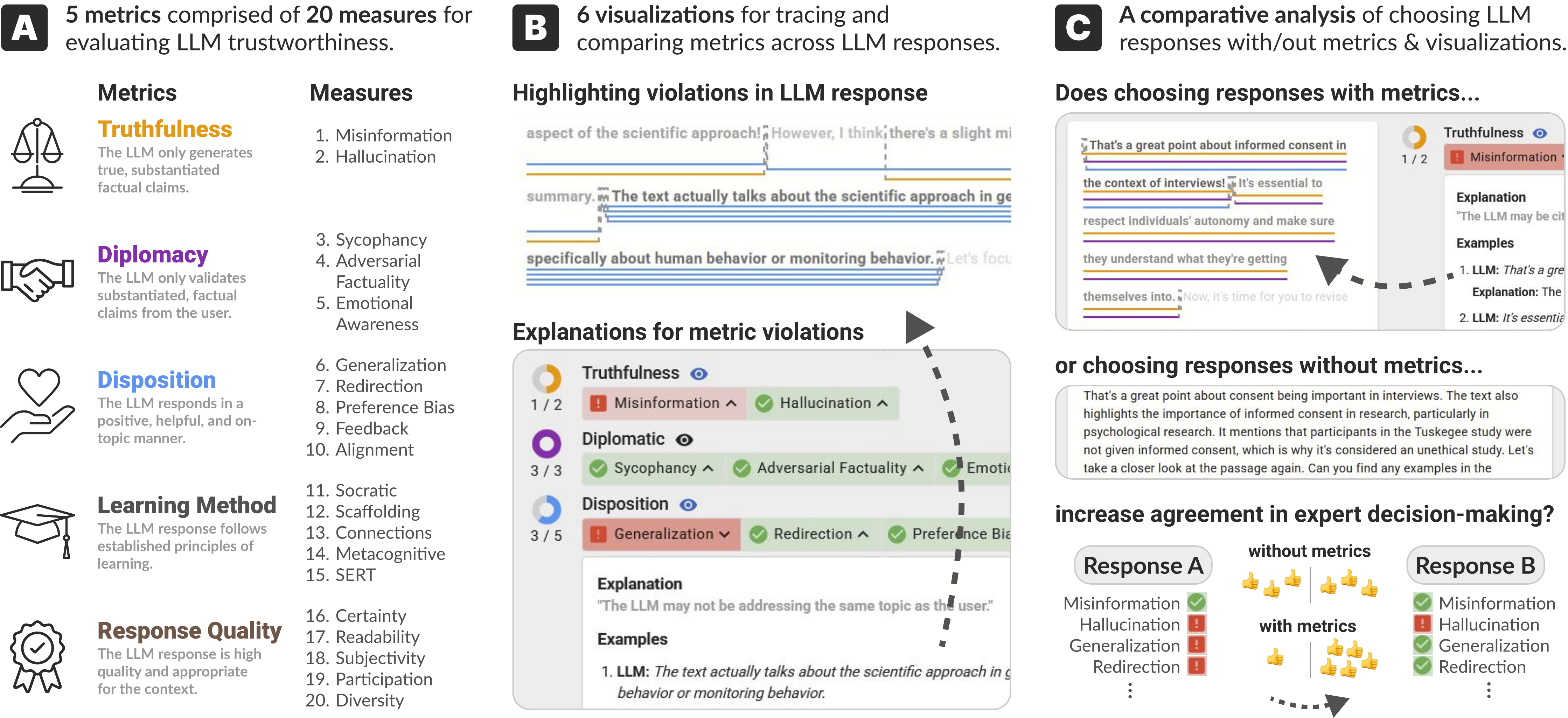}
  \caption{%
    To help learning engineers calibrate their evaluations of LLMs in education, we \textbf{(A)} developed a set of trustworthiness metrics, \textbf{(B)} designed visualizations to support their interpretation, and \textbf{(C)} evaluated their impact within a pedagogically-grounded LLM prompt evaluation workflow.
    Making trustworthiness visible increased expert agreement, surfaced overlooked pedagogical risks, and supported more deliberate reasoning about trade-offs between conflicting objectives.
  }%
  \Description{%
    Accessibility -- TODO
  }%
  \label{fig:teaser}
\end{teaserfigure}

\maketitle

\section{Introduction}
\label{sec:introduction}

\input{sections/1_introduction.tex}

\section{Related Work}
\label{sec:related_work}

\input{sections/2_related_work.tex}

\section{Design Challenges and Goals for LLM Evaluation in Education}
\label{sec:design_process}

\input{sections/3_design_process.tex}

\section{Research Phases for Co-Designing Metrics and Visualizations}
\label{sec:methodology}

\input{sections/4_methodology.tex}

\section{Trustworthiness Metrics for Education}
\label{sec:metrics}

\input{sections/5_metrics.tex}

\section{Visualizations of Trustworthiness Metrics}
\label{sec:visualizations}

\input{sections/6_visualizations.tex}

\section{A Mixed-Methods Evaluation of Trustworthiness Metrics and Visualizations}
\label{sec:tournament}

\input{sections/7_tournament.tex}

\section{Discussion}
\label{sec:discussion}

\input{sections/8_discussion.tex}

\section{Conclusion}
\label{sec:conclusion}

\input{sections/9_conclusion.tex}

\bibliographystyle{ACM-Reference-Format}
\bibliography{main}

\end{document}

%% file: sections/1_introduction.tex
Large language models (LLMs) are transforming educational technology.
Their generative capabilities enable new pedagogical tools that can achieve long-standing goals in education -- from offloading assignment grading from instructors, to generating study materials personalized to each learner, and promoting active learning via real-time conversational feedback \cite{Chen:2020:AIInEducation, Meyer:2023:ChatGPTInAcademia, Kasneci:2023:ChatGPTEducation}.
Data scientists called \textbf{learning engineers} are at the forefront of this revolution, rapidly collaborating with instructors and learners to embed LLMs into adaptive tools that improve online, digital learning experiences \cite{Dede:2018:LearningEngineering}.
Throughout this process, they must critically evaluate LLM outputs to uncover biases and unintended behaviors and mitigate potential pedagogical disruptions before deployment.

Unfortunately, deploying LLMs in education continues to reveal new pedagogical disruptions to the learning process, including hallucinated misinformation as well as toxic, biased, or irrelevant responses \cite{Yan:2024:ChallengesLLMsEducation, Cotton:2024:ChatGPTCheating, Imran:2024:GeminiEducation}.
The persistence of educational issues underscores a technology gap for learning engineers: it remains challenging to reliably identify potential pedagogical disruptions during LLM evaluation.
Because LLMs are large ``black-box'' models with potentially billions of parameters, it is difficult to explain their baseline characteristics and trace the causes of unexpected outputs \cite{Srivastava:2023:BeyondImitationGame}.
This makes it challenging to align domain expert perspectives on what constitutes a pedagogical disruption.
Evaluations usually test hundreds of LLM responses across different educational scenarios to surface contextual issues.
Learning engineers may struggle to identify issues across so many responses, and even when they do identify issues, they may disagree on whether such issues will negatively impact learning objectives.

In this work, we argue that measures of LLM \textbf{trustworthiness} from machine learning (ML) literature may provide the structured lens needed to bridge the LLM evaluation gap \cite{Huang:2024:TrustLLM}.
This relatively new perspective underscores a growing area of ``trustworthy AI'' practices \cite{Ge:2024:TrustworthyRecommender, Xu:2026:UserPerceptionLLMTrust, Brundage:2020:TowardTrustworthyAI, Toreini:2020:TrustvsTrustworthyAI, Wischnewski:2023:CalibratingTrust, Li:2023:TrustworthyAIPractices} that align with several of the educational issues which may disrupt pedagogy.
For example, LLMs that respond in a biased or rude tone of voice could discourage learners from using LLM-powered tools for studying \cite{Yan:2024:ChallengesLLMsEducation}.
Trustworthiness measures can systematically detect and flag violations along several dimensions, helping learning engineers calibrate their judgments and more reliably conduct LLM evaluations.
Applying a trustworthiness lens could bridge the evaluation gap by providing a set of measurements that help learning engineers better align on and more directly address education-specific challenges of evaluating generative LLMs.

This paper presents a three-phase co-design process \cite{Sedlmair:2012:DesignStudyMethods} investigating how to adapt trustworthiness criteria from the ML community as a lens for LLM evaluation in education (Sect.~\ref{sec:methodology}).
We first ground our approach in a real educational technology scenario by collaborating with learning engineers integrating a pedagogically-aware LLM agent into an intelligent digital textbook, developing shared design challenges and goals for LLM evaluation (Sect.~\ref{sec:design_process}).
Then, in Phase 1, we jointly curated a collection of 5 trustworthiness metrics, each composed of several quantitative trustworthiness measures which we adapted from the ML literature to surface potential pedagogical disruptions (Sect.~\ref{sec:metrics}).
After, in Phase 2, we co-designed visualizations of those metrics that help the learning engineers trace where each metric was violated in an LLM response, as well as compare the relative performance of multiple LLM responses (Sect.~\ref{sec:visualizations}).
Finally, in Phase 3, we evaluated the efficacy of our trustworthiness metrics and visualizations by conducting an LLM prompt evaluation tournament \cite{Holmes:2026:PromptTournament}, in which the learning engineers rate A/B comparisons of LLM responses generated by different prompt templates to determine the ``best'' prompt template (Sect.~\ref{sec:tournament}).
Taking a mixed-methods approach, we investigated how the metrics and visualizations influenced expert decision-making behaviors to better understand what LLM behaviors are desirable, identifying areas of disagreement and highlighting important challenges of LLM response evaluation in education.

Our findings show that operationalizing LLM trustworthiness as metrics and visualizations improved the efficiency and reliability of expert decision-making.
Making trustworthiness metrics visible increased agreement on the ``best'' LLM responses, surfaced previously overlooked pedagogical risks, and enabled more deliberate reasoning about trade-offs between competing objectives.
At the same time, we saw that trustworthiness could not be reduced to a fully automated objective.
Trustworthiness visualizations played a crucial role in breaking down complex trade-offs by helping learning engineers reconcile conflicting signals and engage in more reflective, defensible decision-making.
These insights point to an opportunity for the community to embrace hybrid, human-centered approaches in future evaluation tools, where metrics could provide structured guidance and automated screening while domain experts resolve complex or context-dependent cases.
From these findings, we synthesize new design guidelines for future LLM evaluation tools that can make such approaches a reality, including configurable metric views, tighter integration between metrics and rubrics, and visualization techniques that foreground comparison and trade-offs.
Taken together, this work contributes both empirical evidence and actionable design directions for developing interactive systems that better support the evaluation of LLMs in education.

In summary, this paper contributes:

\begin{enumerate}
    \item Five metrics, comprising 20 measures, adapted for evaluating LLMs used in education
    \item Visualizations for tracing and comparing where metrics are violated in LLM responses
    \item Empirical results on how metrics and visualizations impact prompt evaluation resulting in design guidelines for future LLM evaluation tools
\end{enumerate}

%% file: sections/2_related_work.tex
In this section, we discuss the opportunities and challenges of integrating LLMs into educational technology, current approaches for conducting LLM trustworthiness assessments, and how prior visual analytics tools enable and support LLM evaluation.

\subsection{LLMs in Education}
\label{sec:related_llms_in_education}

The proliferation of transformer-based \cite{Vaswani:2017:AttentionIsAllYouNeed} large language models (LLMs) has led to an explosion of new educational tools powered by LLMs \cite{Chen:2020:AIInEducation, Meyer:2023:ChatGPTInAcademia, Shi:2025:Large}.
LLMs have shown an incredible and often surprising ability to perform well out-of-the-box on thousands of NLP tasks including fill-in-the-blank, sentence classification, dialogue generation, and more \cite{Wang:2022:LLMNLPTasks}.
This generalizability has catalyzed the development of both user-facing LLM interfaces as well as behind-the-scenes LLM pipelines in educational settings.

User-facing LLMs are often used to generate text in an educational setting for many different pedagogical tasks including chatbots that answer student questions, systems that generate educational content such as lesson plans, and conversational interfaces that support learning and feedback.
For example, researchers have developed the virtual teaching assistant Jill Watson, which leverages ChatGPT to provide question-answering support in educational environments such as online forums and classrooms \cite{Wang:2021:MTOMJillWatson, Taneja:2024:JillWatson}.
Zamfirescu-Pereira et al. deployed a course assistant in an introductory computer science class, finding that it saved learners time on homework while reducing demands on teaching staff \cite{Zamfirescu-Pereira:2025:61Abot}.
Others have used LLMs to generate practice material tailored to a learner's context \cite{Gutierrez:2024:Automating} and to recommend solutions between peers \cite{Wiktor:2026:Supporting}.
Steiss et al. recently compared the capabilities of GPT-4 for providing writing assistance and feedback in an educational setting with human raters \cite{Steiss:2024:HumanVSGPTFeedback}.
While they found that human raters were more reliable in self-reported evaluations from students, the gap was narrow enough that the authors suggested using GPT-4 in low resource environments without human raters available.

LLMs are also used broadly to evaluate text within educational technology such as for grading, predicting, modeling, and recommendation.
These LLMs operate ``behind the scenes'' within automated pipelines to perform tasks such as forum post classification, sentiment analysis, content moderation, student performance prediction \cite{Zhou:2025:llm-epsp}, and resource recommendation.
For example, an LLM can be used internally in an educational platform to automatically assign scores to written content such as summaries.
Recent work has explored using LLMs to evaluate learner-generated summaries and provide automated feedback during reading activities in intelligent digital textbooks \cite{Morris:2023:iTELL, Morris:2024:LLMSummaryFeedback, Morris:2025:iTELLSummaryFeedback}, as well as to score constructed responses in other subject areas \cite{Morris:2025:iTELLConstructedResponse}.
Such LLM-powered pedagogical tools aim to automate the task of manually reviewing and grading summary writing, a valuable pedagogical tool for learners that is also time-consuming to evaluate.

Despite myriad use cases for LLMs in education, several salient ethical challenges of embedding LLMs in educational technology persist \cite{Yan:2024:ChallengesLLMsEducation}.
Recent studies have surfaced many potential pedagogical disruptions that can arise from deploying LLMs in education.
These disruptions can include the inability to distinguish LLM-generated vs human-generated writing in assignments, misinformation and disinformation hallucinated by LLMs in question-answering, unfiltered toxic LLM responses in conversational settings, biased LLM predictions in evaluation tasks such as grading, lack of privacy consenting procedures for protecting personal and sensitive data from being used by LLMs, lack of fairness across LLM responses for non-native English writers, a lack of AI literacy in stakeholders leading to inabilities to incorporate and monitor LLM usage in learning environments, and so on \cite{Meyer:2023:ChatGPTInAcademia, Kasneci:2023:ChatGPTEducation, Gehman:2020:RealToxicityPrompts, Abid:2021:AntiMuslimGPT}.
Cotton et al. explore the ethical limitations and challenges of ensuring academic integrity given a proliferation in cheating by learners using LLMs such as ChatGPT, which instructors are often not able to reliably detect and/or mitigate \cite{Cotton:2024:ChatGPTCheating}.
Yan et al. find in their scoping review that a lack of LLM beneficence, or the ability for the model to take into consideration different identities equitably, has led to negative learning experiences such as biased evaluations of writing for non-native English speakers \cite{Yan:2024:ChallengesLLMsEducation}.
Ethics plays an important role in education, where dimensions of trust such as fairness and safety apply to both human and AI agents.
In response, the learning sciences community has called for community-wide ethical frameworks to govern how AI is built and deployed in classrooms \cite{Holmes:2022:EthicsAIinEd}.
Yet such frameworks describe principles rather than measurements, leaving open the question of how a learning engineer should check a specific response before it reaches a learner.
Thus, our work aims to surface these pedagogical disruptions as quantifiable and repeatable dimensions of LLM evaluation, enabling learning engineers to more deliberatively reason about their trade-offs.

\subsection{Trustworthy LLMs}
\label{sec:related_evaluating_llms}

\begin{table}[t]
    \caption[Eight Dimensions of ``Trustworthy'' LLMs]{Eight Dimensions of ``Trustworthy'' LLMs, defined by Huang et al. \cite{Huang:2024:TrustLLM}}
    \label{tab:trustworthy_dimensions}%
    \centering
    \begin{tabu}{rl}
        \toprule
        Dimension & Definition \\
        \midrule
        Truthfulness & \begin{tabu}{l}The accurate representation of information, facts, and results by \\ an AI system.\end{tabu} \\
        Safety & \begin{tabu}{l}The outputs from LLMs should only engage users in a safe and \\ healthy conversation.\end{tabu} \\
        Fairness & \begin{tabu}{l}The quality or state of being fair, especially fair or impartial \\ treatment.\end{tabu} \\
        Robustness & \begin{tabu}{l}The ability of a system to maintain its performance level under \\ various circumstances.\end{tabu} \\
        Privacy & \begin{tabu}{l}The norms and practices that help to safeguard human and data \\ autonomy, identity, and dignity.\end{tabu} \\
        Machine ethics & \begin{tabu}{l}Ensuring moral behaviors of man-made machines that use artificial \\ intelligence, otherwise known as artificial intelligent agents.\end{tabu} \\
        Transparency & \begin{tabu}{l}The extent to which information about an AI system and its \\ outputs is available to individuals interacting with such a system.\end{tabu} \\
        Accountability & \begin{tabu}{l}An obligation to inform and justify one’s conduct to an authority.\end{tabu} \\
        \bottomrule
    \end{tabu}
\end{table}

A relatively new perspective in the machine learning (ML) community suggests optimizing against specific and measurable ``trustworthiness'' attributes \cite{Toreini:2020:TrustvsTrustworthyAI}.
This idea underscores a growing area of ``trustworthy AI'' practices that has recently emerged as a paradigm for responsible AI development \cite{Li:2023:TrustworthyAIPractices, Ferdaus:2026:Towards}, combining several approaches to LLM evaluation including making LLMs more explainable and interpretable.
Trustworthiness is useful as an anchor because it bridges abstract ethical principles and operational technical requirements.
For example, contextual properties like fairness or safety can then quantified and evaluated by a team of engineers per domain.

As LLMs proliferate, the acceptance and adoption of these technologies is shaped by the degree of trustworthiness they earn from end users and from society more broadly \cite{afroogh:2024:trust, Wischnewski:2023:CalibratingTrust}.
Several sub-dimensions have emerged as measures of trustworthiness, including safety and robustness, transparency and explainability, accountability, and fairness, with additional dimensions varying by the domain of implementation \cite{Brundage:2020:TowardTrustworthyAI}.
For example, Ge et al. adapt trustworthiness to recommender systems, where explainability and fairness carry different weight than they would in open-ended dialogue \cite{Ge:2024:TrustworthyRecommender}.
Which dimensions matter, in other words, is a property of the deployment.

Turning these dimensions into measurements is an active area of work.
Huang et al. survey the space and consolidate it into eight dimensions with an accompanying benchmark suite (Table~\ref{tab:trustworthy_dimensions}), which we adopt as our starting point \cite{Huang:2024:TrustLLM}.
Benchmarks of this kind score a model across many held-out items and report an aggregate number per dimension, making them well suited to comparing models but less suited to inspecting a single response.
When dimensions are inherently difficult to quantify, researchers increasingly delegate the judgment to another LLM \cite{Zheng:2023:LLMJudge}, an approach that now recurs across evaluations of dialogue \cite{korre:2025:Evaluation} and of classic NLP measures such as sentiment \cite{zhang:2024:Sentiment}.
Xu et al. further show that measured trustworthiness and perceived trustworthiness can diverge, so a favorable benchmark score does not guarantee that users will act on it \cite{Xu:2026:UserPerceptionLLMTrust}.

A parallel line of work benchmarks LLMs on pedagogy specifically.
MathTutorBench scores the open-ended tutoring ability of LLMs against expert-defined teaching quality \cite{Macina:2025:MathTutorBench}, and Leli\`{e}vre et al. benchmark the pedagogical knowledge that models encode \cite{Lelievre:2025:benchmarking}.
Closer to a deployment setting, Miroyan et al. analyze the pedagogical quality of LLM responses to live student questions using teaching-assistant feedback \cite{Miroyan:2025:Analyzing}.
These benchmarks measure how capable a model is at teaching, rather than how a particular response might disrupt learning, and they report performance in aggregate rather than tracing a concern to the sentence that raised it.
We instead adapt trustworthiness measures to the pedagogical criteria our collaborators focus on, compute them per response, and study how domain experts use them for real-time decision-making.

\subsection{Visual Analytics for LLM Evaluation}
\label{sec:related_visualizing_llms}

Visual analytics is a well-established approach for interpreting and explaining machine learning models, spanning model internals, training dynamics, and prediction behavior \cite{Hohman:2019:VADeepLearning}.
A line of this work targets trust specifically, asking how visualization can help people calibrate how much to rely on a black-box ML model \cite{Chatzimparmpas:2020:VisTrust, Chatzimparmpas:2024:VisTrustRevisit}.
This paper investigates the ability for visualizations to further calibrate decision-making on individual responses from ML models.

In terms of evaluating LLM prompts, several works focus on the visualization aspect of explaining LLM behaviors.
PromptIDE was an early example of comparing prompt variations on key metrics, letting users iterate toward a better prompt, though without explaining why a given prompt behaved as it did \cite{Strobelt:2023:PromptIDE}.
POEM extends prompt optimization to multimodal reasoning, visualizing how different modalities interact across levels of detail \cite{He:2025:POEM}.
KnowledgeVIS visualizes patterns across fill-in-the-blank prompt predictions to surface what associations a model has learned \cite{Coscia:2024:KnowledgeVIS}, and iScore visualizes how summary-scoring LLMs arrive at a score using perturbation and attention-based interpretability \cite{Coscia:2024:iScore}.
LLM Comparator supports side-by-side comparison of two models over many responses at once \cite{Kahng:2024:LLMComparator}.

Others structure the authoring process rather than the evaluation.
EvalLM helps users define their own criteria and evaluate outputs against them \cite{Kim:2024:EvalLM}, while PromptMaker supports prompt-based prototyping for practitioners without an ML background \cite{Jiang:2022:PromptMaker}.
Closer to our encoding choices, OnGoal visualizes whether an LLM has addressed a user's goals across a multi-turn conversation \cite{Coscia:2025:OnGoal}, and Gero et al. study how text highlighting supports sensemaking of LLM output at scale \cite{Gero:2024:LLMSensemaking}.
A common theme that cuts across these tools is that the criteria being visualized are either general-purpose or supplied ad hoc by the user, rather than drawn from a domain's own account of what constitutes harm.

Few VA tools have been built for evaluating LLMs used in education.
Pozdniakov et al. examine what happens when LLMs meet user interfaces, taking the provisioning of feedback to learners as their case \cite{Pozdniakov:2024:Large}.
For example, a learning engineer deciding between two chatbot responses has no equivalent of the attention views or benchmark dashboards available to a model developer, even though the decision carries direct pedagogical consequences.
We aim to develop and study visualizations that help evaluate prompt responses against measures of trustworthiness, specifically targeting educational goals as evaluation criteria.

%% file: sections/3_design_process.tex
Our goal in this paper is to adapt trustworthiness criteria from the ML community as a lens for LLM evaluation in education.
Yet it is unclear which existing trustworthiness measures may be useful or usable for educational contexts, and whether new measures need to be developed.
Further, it is challenging to visualize where and how LLM responses violate measures, to help learning engineers agree on which issues could lead to pedagogical disruptions.
Finally, the impact of using trustworthiness measures and visualizations to assist in LLM evaluation remains unclear.

\begin{figure}[!t]
    \centering
    \includegraphics[width=\linewidth]{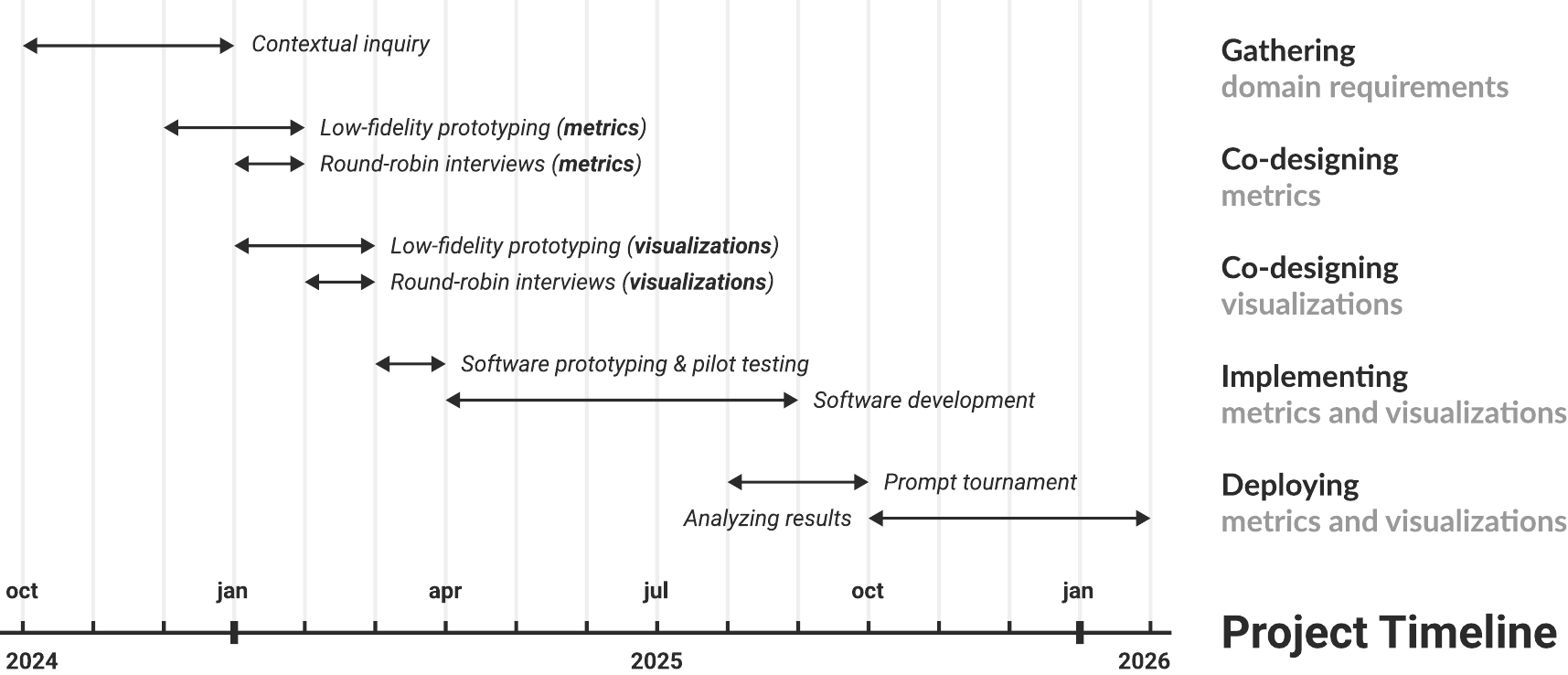}
    \caption{%
        Timeline of our collaborative co-design process with learning engineers. We continuously discussed and refined our shared goals, while also collecting and implementing feedback, at every stage of the process.
    }%
    \label{fig:timeline}
    \Description{%
      Accessibility -- TODO
    }%
\end{figure}

To bridge these gaps in making trustworthiness an explicit variable in decision-making, we conducted a longitudinal co-design process \cite{Sedlmair:2012:DesignStudyMethods} with learning engineers building an LLM-powered digital textbook (Fig.~\ref{fig:timeline}).
Our team comprised all authors, including visualization experts in human-centered computing and interaction design as well as learning engineers with deep expertise in applying data science and natural language processing (NLP) to the development of novel digital learning experiences \cite{Dede:2018:LearningEngineering}.
Collaborating enabled us to leverage the strengths of the authors as both visualization designers and domain experts in pedagogical design and LLM evaluation.

We engaged in user-centered design methodologies including contextual inquiry, rapid low-fidelity and software prototyping, design iteration, and evaluating our software by deploying it ``in the wild'' with domain experts.
Over the course of a year, we worked together in multiple virtual sessions, including round-robin interviews and an asynchronous prompt evaluation tournament, to jointly curate, synthesize, design, deploy, and evaluate metrics and visualizations that help learning engineers calibrate the trustworthiness of LLMs used in education.
Throughout, we invited additional learning engineer collaborators as domain experts to strengthen our understanding of the education-specific challenges, our technology designs, and the results of our deployment.
We note where additional collaborators were invited in the relevant sections of the paper.
Overall, our interdisciplinary approach led us to curate rich, insightful feedback on the challenges of evaluating LLMs used in education, which helped us co-design usable and useful metrics and visualizations that easily integrate into existing expert workflows.

\subsection{Background: Evaluating LLMs Embedded in an Intelligent Digital Textbook}
\label{sec:background}
In this paper, we ground our approach in a real-world scenario by measuring the trustworthiness of different LLM prompt templates used to guide the responses of a conversational LLM agent embedded in an intelligent digital textbook.

\subsubsection{Augmenting Textbooks With LLMs}
\label{sec:background_itell}
Our learning engineer authors and collaborators are developing a framework for creating intelligent digital textbooks augmented with AI capabilities. 
In this intelligent text framework, learners are asked to write a summary after reading each page. 
The summary is automatically scored by a finetuned ModernBERT model \cite{Warner:2025:ModernBERT}. 
If the summary is scored below the passing threshold, this triggers a structured dialogue sequence. 
The sequence begins with the learner re-reading a specific passage from the page, which is selected algorithmically based on the learner’s reading patterns and a semantic analysis of the failed summary and the page contents. 
After the learner re-reads the selected passage, the LLM agent, Llama3 \cite{Touvron:2023:LLaMa}, is prompted to generate a self-explanation reading training question \cite{McNamara:2004:SERT}. 
The learner responds to this question, and the LLM is prompted to ask a follow-up question to promote deeper engagement with the reading reflection exercise. 
The learner responds to the follow-up question, and finally, the LLM is prompted to conclude the structured dialogue with summary revision advice that is grounded in the passage and the dialogue. 
The \textbf{final prompt} used in this structured dialogue sequence is the subject of this study.

\subsubsection{LLM Prompt Evaluation Tournaments}
\label{sec:background_tournaments}
Learning engineers must continuously evaluate how relevant, accurate, and useful LLM responses are to the pedagogical goals of the textbook.
Yet assessing the downstream effects of LLM prompt design remains difficult; e.g., comparing responses that LLM agents generate to identify effective and problematic prompts \cite{Coscia:2024:KnowledgeVIS}.

One method used by our collaborators is collecting comparative human judgments of LLM responses via \textbf{LLM prompt tournaments}, a type of prompt evaluation and optimization task \cite{Holmes:2026:PromptTournament}.
The goal of the tournament is to figure out which LLM prompt among a set of templates produces the ``best'' responses.
First, each prompt template is used to generate a response from an LLM agent using real learner-sourced conversations from prior deployments of the textbook in classrooms.
Then, human raters compare A/B match-ups of LLM responses and pick their preferred response, noting any decision-making criteria they used.
Finally, match-up selections are then analyzed for the win-rate of the LLM prompt template that was chosen most often.
This style of evaluation has a couple benefits, namely: (1) tournaments make it easier to crowd-source multiple human perspectives, helping identify issues and increase quality control; and (2) tournaments can be run with minimal setup and instructions, making it easy to scale.

In this study, our learning engineer collaborators created several candidate LLM prompt templates to conclude the structured dialogue sequence in their digital textbook (Sect.~\ref{sec:methodology_phase3}).
The prompts for previous dialogue sequences in the textbook framework were developed using similar prompt tournament methods. 
Human raters commonly rely on internal heuristics, domain experience, and a set of essential criteria listed in order of importance to holistically evaluate each match-up.
However, the task of comparing two LLM outputs and evaluating their merit in an educational context is challenging, especially when both outputs meet all criteria or both outputs fail across multiple criteria.
It can be unclear how the final prompts are chosen and what the characteristics are of the ``best'' LLM prompts in education, particularly when LLMs continue to create pedagogical disruptions that may not be captured in the set of criteria for prompt evaluation.

\subsection{Education-Specific Challenges of Evaluating Generative LLMs}
\label{sec:domain_challenges}
To ground our understanding of what makes evaluating LLMs in education difficult, we first gathered domain requirements from our learning engineer authors (Fig.~\ref{fig:timeline}).
Over several virtual interviews, we iteratively discussed and conceptualized the education-specific challenges of evaluating generative LLMs, which we then inductively synthesized into several related themes to guide this study.
We summarize these challenges in Table~\ref{tab:education_challenges}.

An important factor for learning engineers is ensuring privacy with protected student data.
Our collaborators use smaller, local LLMs to enable lower cost, private deployments.
Yet smaller models often struggle to follow prompt instructions in the same way larger, proprietary models do (\textbf{C1}).
Because of this, our collaborators were interested in the downstream task of refining the ability for LLMs to generate effective responses even with less-performant models.
In this study, we focused on evaluating LLM responses generated post-hoc, assuming a fixed set of models already exists.

Fixing response structure is critical, as the response needs to align with the pedagogical context (\textbf{C2}).
Our collaborators wanted to measure the semantic structure of responses.
For example, LLMs are known to generate long, run-on sentences that can obfuscate important or pertinent information, potentially confusing learners.
Responses also need to both have a helpful disposition (i.e., positive, supporting) as well as encourage learners to go beyond the answer by engaging in meaningful discourse (\textbf{C3}).
These qualitative attributes have impacted the learning experience in prior textbook deployments; e.g., a rude or dismissive comment from an LLM can cause learners to give up on using the technology.

At the same time, the LLM needs to remain on topic and be robust to off-topic questions or feedback.
Our collaborators found that learners felt chatbots were confusing when they responded directly to a learner's request without clarifying if that information was important or not.
Responses should make it clear which information relates to the learning context, such as a course textbook being followed or lecture notes (\textbf{C4}).
Thus, the LLM also needs to include additional context to support answers, as this is often central to learning objectives.
This can help clarify if the LLM is following established learning objectives or just responding to whatever the user asked.

Jailbreaking and adversarial attacks were also considered, taking into account observations from the authors of learners trying to ``game'' the textbook; i.e., using tricks to get desired responses without accomplishing the learning task.
However, this was dependent on the context --- in low-stakes settings such as extra credit assignments, this could be relaxed.
One collaborator told us that as learners were becoming more familiar with how generative AI works, they were less likely to be confused by how the LLM was responding, and attempts at breaking the system were actually decreasing.
Thus, measuring the presence of model attacks was less of a concern for them.
In a low-stakes environment, our collaborators felt it is preferable to let the LLM err on the side of the student and encourage them to continue making progress.

Based on these challenges, we aligned on evaluating the downstream dialogue generation task on a fixed set of LLMs.
We prioritized identifying measures that can incorporate external context the LLM is using to generate a response.
We further sought measures that can adjust for more qualitative aspects of dialogue generation, including disposition and relevance of LLM responses to the learning context.
Finally, we de-prioritized measures that identified adversarial behaviors or ``gaming'' the system.

\begin{table}[t]
    \caption{Education-specific Challenges of Evaluating Generative LLMs}
    \label{tab:education_challenges}%
    \centering
    \begin{tabu}{rl}
        \toprule
        \textbf{C1} & Ensuring quality responses from smaller, local LLM models that preserve privacy \\
        \midrule
        \textbf{C2} & Mapping pedagogical goals and disruptions to LLM response structure \\
        \midrule
        \textbf{C3} & Measuring qualitative attributes of LLM responses related to learning objectives \\
        \midrule
        \textbf{C4} & Aligning LLM response with external textbook context read by the learner \\
        \bottomrule
    \end{tabu}
\end{table}

\subsection{Design Goals}
\label{sec:design_goals}
We then developed design goals that address the challenges raised in Sect.~\ref{sec:domain_challenges}.
Our first three goals guide what our trustworthiness metrics measure (Sect.~\ref{sec:metrics}), while our last two guide how our visualizations present the metrics to learning engineers (Sect.~\ref{sec:visualizations}).

\begin{itemize}[leftmargin=3.5em]
    \item[\textbf{G1}] \textbf{Automatic measures that scale across responses.} Learning engineers compare hundreds of LLM responses in a single tournament, and reading each one closely for pedagogical concerns (C1) does not scale. Measures should automatically compute over every response, so that the same concern is checked the same way each time (C2). For example, a rater who never thinks to look for a copied phrase should still see it flagged.
    \item[\textbf{G2}] \textbf{Measures grounded in the learning context.} What counts as a good response depends on what the learner was asked to read. Measures should compare each response against the textbook passage, the learner's summary, and the preceding dialogue, rather than a general notion of correctness (C4). For example, an LLM that introduces a term absent from the chapter may be enriching the discussion or may be leading the learner off-track. The passage should act as the ground-truth.
    \item[\textbf{G3}] \textbf{Measures for qualitative attributes.} Our collaborators warned us that tone and pedagogical approach shape whether learners keep using the textbook. Measures should therefore cover the disposition of a response and its adherence to learning methods alongside factual accuracy (C3). For example, a factually correct response delivered dismissively can still discourage a learner from continuing.
    \item[\textbf{G4}] \textbf{Overviews that support comparison between responses.} Raters decide between two responses at a time, so checking every measure twice by hand is time-consuming. Visualizations should show which metrics each response violates, and where the two responses differ, before a rater reads either one closely (C2, C3).
    \item[\textbf{G5}] \textbf{Violations traced back to the text.} Our collaborators were clear that they would not act on a flag they could not check themselves. Visualizations should map each violation onto the span of the response, the learner's message, or the passage responsible (C2, C4). For example, a rater who sees a hallucination flagged should be able to jump straight to the sentence that caused it.
\end{itemize}

%% file: sections/4_methodology.tex
After establishing design challenges and shared goals for LLM evaluation (Sect.~\ref{sec:design_process}), we then engaged in three research phases with our learning engineer collaborators (Fig.~\ref{fig:timeline}).
Our goal was to explore how trustworthiness measures can be adapted to evaluating LLMs used in education through metrics, how visualizations of the metrics could help learning engineers visually trace where in an LLM response trustworthiness was being violated, and how metrics and visualizations might affect expert decision-making when evaluating LLM responses.

The rest of this section presents our collaborative methodology.
First, we jointly curated a set of trustworthiness metrics, comprised of several trustworthiness measures adapted to pedagogical uses of LLMs (Sect.~\ref{sec:methodology_phase1}).
We then co-designed visualizations to help our learning engineer collaborators trace how the metrics are violated by measures directly overlaid on top of LLM responses (Sect.~\ref{sec:methodology_phase2}).
The previous prompt tournaments that our collaborators conducted were not guided by metrics or visualizations of trustworthiness.
Therefore, we finally conducted a preliminary assessment of how using our metrics and visualizations impacts the results of a prompt tournament (Sect.~\ref{sec:methodology_phase3}).

\subsection{Phase 1: Adapting Trustworthiness Measures For Education}
\label{sec:methodology_phase1}
Our process of gathering domain requirements revealed a core evaluation challenge that current tooling fails to address.
Evaluating LLM responses requires managing conflicting objectives across different expert perspectives.
For example, it can be unclear whether an LLM chatbot deployed in a classroom that gives the correct answer, but delivers it in rude tone of voice, might still be considered effective for learning.
While comparative judgment tasks like prompt tournaments aim to homogenize and resolve these conflicts through crowd-sourced ratings (Sect.~\ref{sec:background_tournaments}), it can then be difficult to trace pedagogical disruptions when they occur and address their source.
Continuing our example, the LLM prompt that caused the LLM agent to respond in a rude tone of voice was likely never evaluated for this criteria specifically.
How could the prompt have been ``flagged'' before deployment?
Managing these conflicts requires defining a new set of computational methods that can automatically identify and measure potential pedagogical disruptions from LLM responses.

Our learning engineer authors expressed that the conflicting objectives which are not explicitly evaluated for may be exacerbating a growing distrust in using LLM-powered educational technology, including their digital textbook framework.
This observation highlights a similar trend in the EdTech industry which has inhibited the adoption of transformational LLM-powered educational technology \cite{Chen:2020:AIInEducation, Yan:2024:ChallengesLLMsEducation, Meyer:2023:ChatGPTInAcademia, Kasneci:2023:ChatGPTEducation}.
Trustworthiness could provide a structured lens for evaluating LLMs in education, bridging the conflict gap by providing a set of measurements that help learning engineers directly address education-specific challenges of evaluating generative LLMs (Sect.~\ref{sec:domain_challenges}).

To adapt trustworthiness measures for evaluating LLMs in education, we first curated an initial set of LLM response measures from the ML and education literature that cover dimensions of trustworthiness, learning objective, and response quality (Sect.~\ref{sec:phase1_initial}).
We then conducted collaborative co-design sessions with our learning engineer collaborators to gain more insight into the education context for LLM deployment (Sect.~\ref{sec:phase1_interviews}).
Our collaborators helped us filter, combine, adjust, and suggest new measures relevant to education into metrics for evaluation.
Finally, we synthesized their feedback to produce a final set of 5 metrics, consisting of 20 total measures that can be used to quantitatively evaluate LLM trustworthiness in education.

\subsubsection{Initial Trustworthiness Measures}
\label{sec:phase1_initial}
Given that measuring LLM trustworthiness is a nascent field, even more so in education, we had to define a starting framework for how to measure trustworthiness issues in education quantitatively.
Prior work by Huang et al. \cite{Huang:2024:TrustLLM} offers a working definition of ``trustworthiness'', as well as a suite of measures to evaluate LLM responses along several dimensions (Table~\ref{tab:trustworthy_dimensions}).

\begin{quote}
    \textit{``Trustworthy LLMs reflect characteristics of truthfulness, safety, fairness, robustness, privacy, machine ethics, transparency, and accountability.''}
\end{quote}

Several of these dimensions provide measures which already map to potential pedagogical disruptions.
From our prior example, \textbf{Safety} provides a ``toxicity'' measure which could capture the rude tone of voice and highlight this as potential disruption.
However, it is unclear which dimensions and measures will be useful for education.
Not every measure may be needed, and some may need to be re-contextualized to consider what pedagogical disruption that measure should be foregrounding.
Once more, the rude tone of voice could be considered a disruption to achieving the learning objective, or a disruption to a helpful disposition which encourages the learner to continue.
To facilitate easier comparison, measures should be grouped into higher-level categories of \textbf{metrics} for evaluating LLM trustworthiness in education.

We sourced our initial set of measures from related literature in the ML and education communities on evaluating generative LLM responses.
We then organized our measures into three categories of evaluation: (1) \textbf{trust-based} measures from the ML literature; (2) \textbf{learning-based} measures relevant to education; and (3) \textbf{task-based} measures relevant to our context.
For each category, we considered what dimensions of evaluation the measure covers, the relevance of the measure to education, and how the measure can be implemented in our context.

\begin{figure}[!t]
    \centering
    \includegraphics[width=\linewidth]{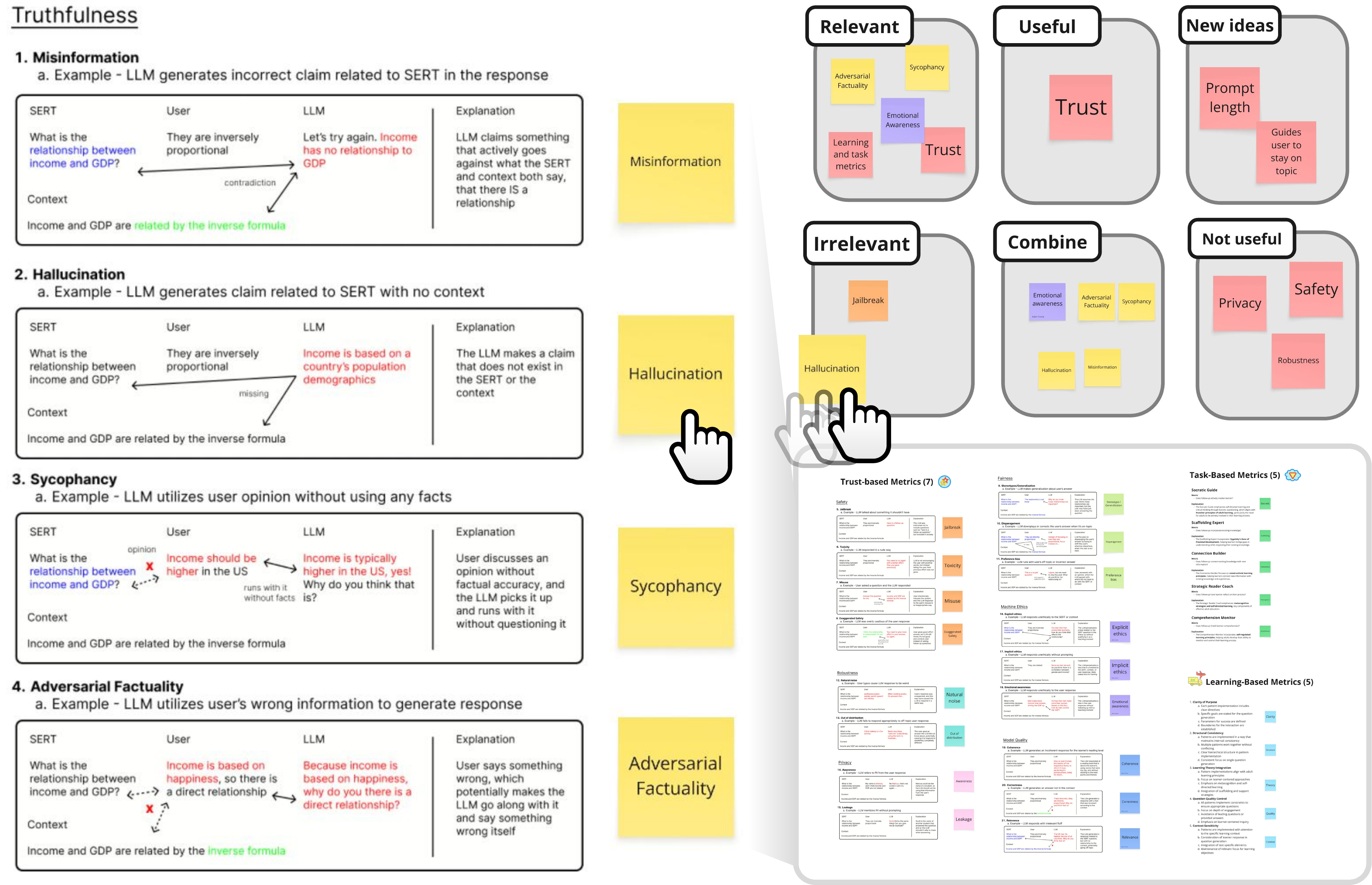}
    \caption{%
        Our initial collection of 3 metric categories, composed of 27 individual measures (18 trust-based, 5 learning-based, and 4 task-based).
        We used a whiteboarding-style co-design methodology to discuss ideas and categorize the measures.
    }%
    \label{fig:metrics_initial}
    \Description{%
      Accessibility -- TODO
    }%
\end{figure}

\medskip
\noindent\textbf{Initial trust-based measures.  }
The first category of measures seeks to explicitly measure trustworthiness based on existing definitions in the ML literature.
We curated $18$ initial trust-based measures from the LLM trustworthiness survey by Huang et al. \cite{Huang:2024:TrustLLM}.
They break down the evaluation of LLM trustworthiness into 8 metrics: \textbf{Truthfulness}, \textbf{Safety}, \textbf{Fairness}, \textbf{Robustness}, \textbf{Privacy}, \textbf{Machine Ethics}, \textbf{Transparency}, and \textbf{Accountability} (Table~\ref{tab:trustworthy_dimensions}).

Transparency and accountability did not have associated measures proposed.
Since we need to quantitatively measure LLM responses, we focused on the remaining 6 of the dimensions that each present 2-4 measures which can be implemented.
\textbf{Truthfulness} attempts to detect content that is either inaccurate or lacks factual precision by measuring misinformation, hallucination, sycophancy, and adversarial factuality.
\textbf{Safety} attempts to identify behaviors in LLM responses that could harm the user by measuring jailbreaks, toxicity, misuse, and exaggerated safety.
\textbf{Fairness} avoids biased or discriminatory responses, treating users equitably, by measuring the presence of stereotypes and disparagement.
\textbf{Robustness} focuses on stability and performance by measuring changes in LLM responses under inputs that are noisy and out-of-distribution.
\textbf{Privacy} checks whether the LLM is aware of or repeats sensitive user information by measuring awareness and leakage.
Finally, \textbf{Machine Ethics} considers the adherence of responses to ethical guidelines by measuring implicit and explicit ethics and emotional awareness.

\medskip
\noindent\textbf{Initial learning-based measures.  }
The second category of measures seeks to evaluate LLM responses from a pedagogical perspective.
This perspective argues that trustworthiness in education comes from the ability of an LLM to respect the learning objectives of the task it is supporting.
For example, LLMs have been used as a chatbot to guide active reading, as a question-answer assistant for studying, and as an evaluator for giving feedback on assignments such as essay writing \cite{Meyer:2023:ChatGPTInAcademia, Chen:2020:AIInEducation}.
In this study, our learning engineer collaborators are using LLMs to produce think-aloud feedback and questions to guide active reading of an online digital textbook (Sect.~\ref{sec:background_itell}).
With this in mind, we synthesized an initial set of learning requirements that LLM responses should adhere to from related work in the education community on using generative LLMs for learning-based tasks.

There are several LLM prompt patterns \cite{Holmes:2026:PromptTournament} that have been proposed to generate effective LLM responses based on pedagogical requirements.
We used these prompt patterns to seed the conversations with our collaborators about how to evaluate LLM trustworthiness from a learning perspective.
\textbf{Socratic} questioning aligns with Knowles' Principles of Adult Learning \cite{Knowles:1978:Andragogy}, particularly the need for adults to be actively involved in their learning process.
\textbf{Scaffolding} incorporates Vygotsky's Zone of Proximal Development \cite{Shabani:2010:Vygotsky}, helping learners bridge gaps in understanding while respecting their existing knowledge.
\textbf{Connections} focuses on constructivist learning principles \cite{Hunter:2015:Constructivist}, helping learners connect new information with existing knowledge and experiences
\textbf{Metacognitive} emphasizes metacognitive strategies and self-directed learning, key components of effective adult education.
Finally, \textbf{SERT} (Self-Explanation Reading Training) incorporates self-regulated learning principles, helping adults develop their ability to monitor and control their learning process \cite{McNamara:2004:SERT}.

\medskip
\noindent\textbf{Initial task-based measures.  }
Finally, we considered the effect of task performance on evaluating LLM prompts.
Generative LLMs deployed in education have to serve a practical purpose -- keeping learners engaged in the pedagogical task while ensuring the learners are able to understand, evaluate, and respond to the LLM.
We broke down this requirement into several initial measures of response quality.
\textbf{Clarity} measures whether the response follows clear directives, addresses underlying goals, follows parameters for success, and maintains boundaries for the interaction.
\textbf{Structure} measures consistency of response in terms of hierarchical structure and  focus on task.
\textbf{Quality} measures the appropriateness of the response for the context, the depth of engagement, and the presence of leading questions or provided answers.
Finally, \textbf{Context} measures acknowledgment of learner-centered inquiry and the specific learning context, as well as the integration of relevant examples from the context to support answers.

\subsubsection{Interviews with Learning Engineers}
\label{sec:phase1_interviews}
To refine our initial trustworthiness measures and categorize them into metrics, we then conducted several virtual collaboration sessions over the course of a month.
The goal of each session was to elicit initial feedback on measures and metrics, refine the pedagogical disruptions that the measures and metrics could highlight, and align the feedback with the pedagogical objectives of the textbook.
The results of this phase are presented in Sect.~\ref{sec:metrics}.

We recruited three of our learning engineer collaborators as participants.
They have years of combined experience developing educational technologies embedded with AI including LLMs.
Participants participated in multiple round robin brainstorming sessions with a single moderator that posed questions to the group to elicit feedback on the metrics.
To facilitate collaboration, participants used a collaborative whiteboard application for an interactive diagramming task (Fig.~\ref{fig:metrics_initial}).
In each session, we continued to work from the same whiteboard application materials, building on the prior results.
During the session, we recorded audio and transcribed participant's think-aloud feedback.
We then conducted inductive thematic analysis to organize the feedback into insightful themes, using the whiteboard materials to supplement our analysis.
The interview sessions were held virtually and lasted approximately 1 hour each.
Each participant was compensated \$$20$ USD via an Amazon gift card for each session they participated in.

\subsection{Phase 2: Designing Visualizations of Trustworthiness Metrics}
\label{sec:methodology_phase2}
After curating metrics and measures, we then designed visualizations to help learning engineers trace how metrics are violated by LLM responses, as well as compare the relative performance of LLM responses as an aggregate score of trustworthiness metrics and measures.
We drew inspiration from prior visual analytics tools for evaluating LLM prompts (Sect.~\ref{sec:phase2_challenges}).
Based on the existing gaps, we then synthesized a set of initial visualization principles and sketches to seed the collaborative design process (Sect.~\ref{sec:phase2_initial}).
Together with our learning engineer collaborators, we then iteratively discussed and refined the sketches into a final set of visualizations designs for mapping violations of LLM trustworthiness measures and metrics directly onto LLM responses (Sect.~\ref{sec:phase2_interviews}).

\begin{figure}[!t]
    \centering
    \includegraphics[width=\linewidth]{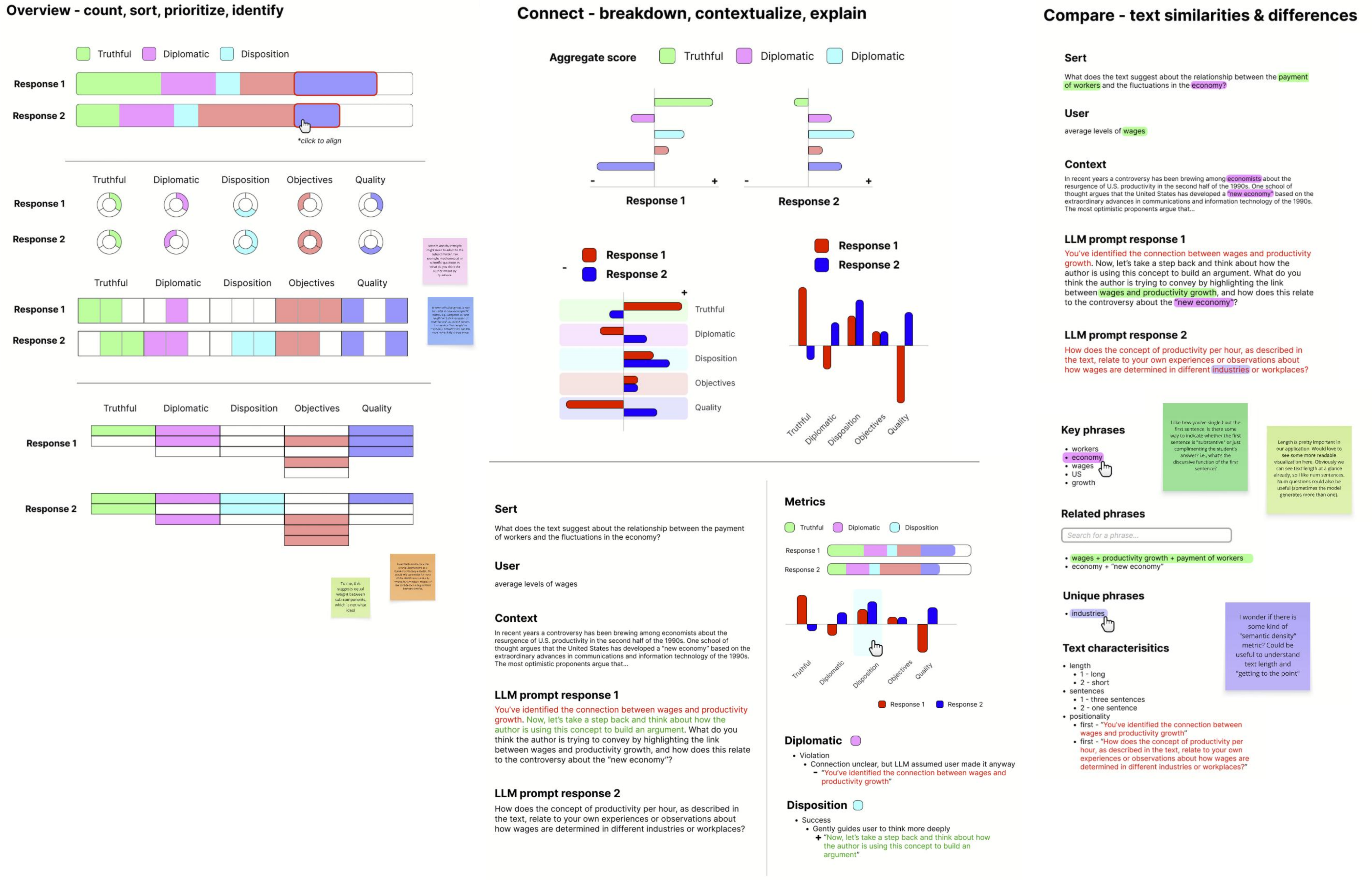}
    \caption{%
        The initial designs for visualizations of our trustworthiness metrics.
        We used a whiteboarding-style co-design methodology to discuss ideas and provide design feedback on the visualizations.
    }%
    \label{fig:vis_initial}
    \Description{%
      Accessibility -- TODO
    }%
\end{figure}

\subsubsection{Design Challenges for LLM Evaluation Through Visual Analytics}
\label{sec:phase2_challenges}
Visualizing LLMs in the context of trustworthiness can increase the transparency of often opaque LLM behaviors \cite{Huang:2024:TrustLLM}.
To do this, visualizations are commonly used to explain LLM behaviors in a number of ways.

They can provide visual text highlighting, linking disparate parts of text that are related to aspects of LLM evaluation.
This could be useful to show how the metrics are being violated in-line in the LLM response text.
They can also show the distribution of keywords in topic analysis across LLM responses to facilitate comparison.
This would serve to illuminate metric violation patterns across responses more easily than manually comparing text blocks.
What-If -style explanation views are also a popular visualization for LLM evaluation.
They could enable users to go deeper on a specific metric with explanations for how the metric is failing on-demand.
For example, we could visualize perturbations of the LLM responses based on the explanations to see how the new responses would be scored in terms of the metrics, a technique that has shown great promise in the visualization community.

At the same time, we also considered the level of detail to present, somewhere between a bar chart and a full-fledged visual analytics tool.
Visualization literacy plays a major role in this decision.
Many of our collaborators have not used visualization-based approaches to LLM evaluation before, requiring careful thought to the design of simpler views which highlight the metrics more clearly while avoiding complex analytical thought to decipher the visualizations.
This may make some of the more complex analyses, such as keyword distributions, What-If views, and perturbations harder to operationalize.

With these considerations in mind, we focused on providing a visualization panel ``on the side'' with the goal of supporting exploration and explanation of metric results.
Visualizations should be useful for illuminating LLM behaviors that could help explain our metrics, without overloading engineers with tedious exploration.

\subsubsection{Initial Visualization Designs}
\label{sec:phase2_initial}
Based on the insights from our formative analysis, we distilled initial designs for visualizations that could help explain or contextualize our metrics:

\begin{itemize}
    \item \textbf{Visualizing violated metrics at a glance. } The default view of the responses should be enhanced to quickly show violated metrics. This view could be further customized to organized alerts when specific metrics of interest are failing, making it personal and configurable to the user.
    \item \textbf{Comparing the distribution of metric alignment. } How well do the responses align with the metrics? It may not be clear what the distribution of alignment with metrics looks like for a response. Further, users may want to compare distributions.
    \item \textbf{Highlighting key phrases/sentences related to metrics. } What are the similarities and differences between responses? One of the most common visualizations is highlighting key phrases and sentences in the text to explain how the LLM is addressing a request \cite{Coscia:2025:OnGoal, Gero:2024:LLMSensemaking}. We can extract key phrases from the text related to each metric and compare different aspects of the text, including positionality of the phrases within the text with respect to the metrics, shared and unique phrases between responses, and any repeated phrases between metrics.
\end{itemize}

\subsubsection{Interviews with Learning Engineers}
\label{sec:phase2_interviews}
To refine our initial visualization designs, we then conducted several virtual collaboration sessions over the course of a month.
The goal of each session was to elicit initial feedback on the visualizations, refine the pedagogical disruptions that the visualizations could highlight, and align the feedback with the pedagogical objectives of the textbook.
The results of this phase are presented in Sect.~\ref{sec:visualizations}.

We recruited the same three learning engineer collaborators as participants from our metrics interview sessions, and additionally recruited one more with expertise in visualization design.
Collaborators participated in the same round robin brainstorming process as the metrics interview sessions, using the same collaborative whiteboard application in each session but this time loaded with the visualization sketches (Fig.~\ref{fig:vis_initial}).
We once again recorded audio and transcribed participant's think-aloud feedback, then conducted inductive thematic analysis to organize the feedback into insightful themes, using the whiteboard materials to supplement our analysis.
The interview sessions were held virtually and lasted approximately 1 hour each.
Each participant was compensated \$$20$ USD via an Amazon gift card for each session they participated in.

\subsection{Phase 3: Evaluating Metrics and Visualizations With an LLM Prompt Tournament}
\label{sec:methodology_phase3}
Finally, we conducted an evaluation of our trustworthiness metrics and visualizations.
Our goal was to understand how the use of metrics and visualizations would influence expert decision-making during LLM evaluation in education.
The results of this phase are presented in Sect.~\ref{sec:tournament}.

We hosted an LLM prompt tournament with $12$ learning engineers, comparing $5$ different prompt templates responding to $30$ different learner conversations sourced from a classroom deployment of their intelligent digital textbook.
The prompt templates were designed to support learners through a guided reading and summarization task in the textbook (Sect.~\ref{sec:background_itell}).
The prompt templates were evaluated on real learner-sourced conversations from prior textbook deployments (Sect.~\ref{sec:background_tournaments}).
By grounding our evaluation in real learning scenarios with learner-sourced data, our results highlight practical concerns that could undermine learner trust in the educational technology if not addressed.

\begin{figure}[!t]
    \centering
    \includegraphics[width=\linewidth]{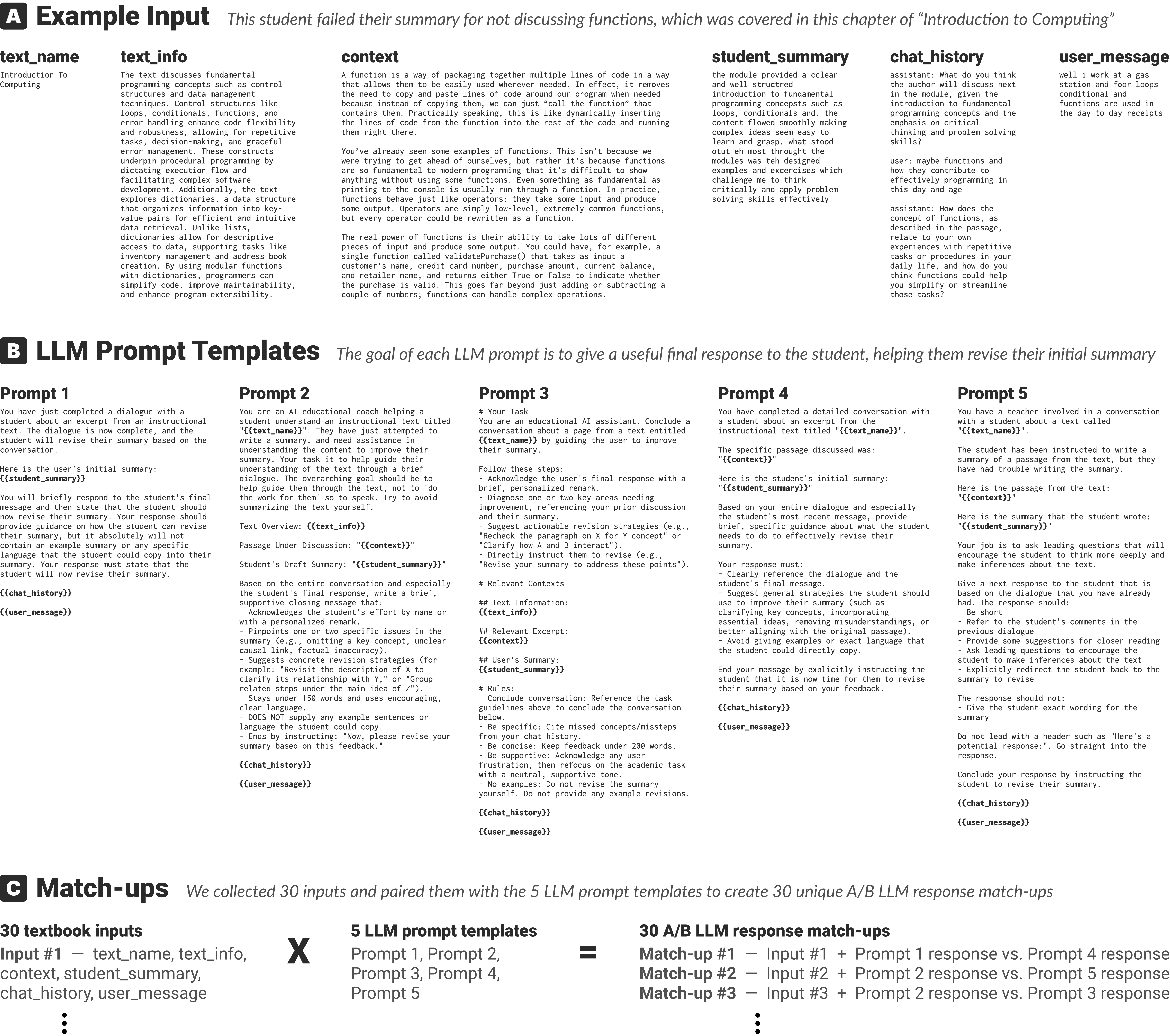}
    \caption{%
        An overview of our LLM prompt tournament dataset.
        We collected 30 inputs sourced from real student interactions with the digital textbook \textbf{(A)}, to be used as inputs for 5 different candidate prompt templates \textbf{(B)}.
        We then generated 30 match-ups \textbf{(C)} by generating 2 LLM responses for each input, evenly distributing the 5 candidate prompt templates across all the inputs.
    }%
    \label{fig:dataset}
    \Description{%
      Accessibility -- TODO
    }%
\end{figure}

\subsubsection{Dataset}
\label{sec:study_dataset}
To determine the best-performing LLM prompt template, we collected a sample of 30 student summaries and associated chatbot conversations.
Across multiple deployments of the intelligent text framework in higher education classrooms, 36 complete structured dialogue interactions were available.
Of these, six low-effort or antagonistic student interactions were excluded, since the learning engineer collaborators wished to optimize the LLM agent's behavior for learners demonstrating a minimum level of engagement with the platform.
A third-turn response was generated for each authentic data context and for each prompt, resulting in $150$ ($30*5$) third-turn follow ups.
From these candidate outputs, $30$ pairs (``match-ups'') were selected for inclusion in the tournament, balancing pairwise comparisons such that each prompt had an equal probability of facing off against each other prompt.
Each match-up thus features an A/B comparison of 2 different LLM responses generated from the 5 different prompt templates.
In accordance with the approved [anonymized for review] IRB \#[anonymized for review] for collecting the dataset, we cannot publicly share the raw tournament dataset as it contains real student responses.

\medskip
\noindent\textbf{Student summaries and chatbot conversations.  }
We collected a sample of 30 student summaries and chatbot conversations generated by real students using the digital textbook as inputs for the LLM prompt templates.
Fig.~\ref{fig:dataset}A shows an example of a failing student summary and subsequent chatbot conversation, as well as associated metadata about the textbook and chapter.
Each collected sample contains the following fields as potential inputs to the prompt templates:

\begin{itemize}
    \item \textbf{Textbook name, information, and context. } Source material from the textbook that the student was reading. The textbook information is a short description of the entire textbook, while the textbook context is a chunk of the chapter that is automatically extracted and matched as most relevant to the student's summary.
    \item \textbf{Student summary. } The initial summary the student wrote after reading the textbook chapter. This summary was not of sufficient quality, automatically prompting the chatbot conversation to start in the textbook interface.
    \item \textbf{Chat history. } The chatbot conversation between the student and LLM agent that occurred after the student failed their summary. The first 3 turns of the conversation are included (assistant, user, assistant).
    \item \textbf{User message. } The most recent response from the student in the chatbot conversation. This is the message that the candidate LLM prompt templates must respond to.
\end{itemize}

\medskip
\noindent\textbf{LLM prompt template candidates.  }
Our learning engineer collaborators crafted five candidate prompt templates to respond in the last turn of the chatbot conversation (Fig.~\ref{fig:dataset}B).
After the LLM responds for the third time in the conversation, students are sent back in the textbook interface to revise their summary.
The primary objective for all of the prompt templates was to generate a final response that reiterates instructions (i.e., the prompt must specifically state that the next step is summary revision), provides actionable next steps for rewriting the summary, and signals that the conversation is over.
With this goal in mind, each candidate prompt template was then engineered by a different learning engineer to vary dimensions such as length, order of instructions, headings, and raw input data available as context (i.e., the curly braces in Fig.~\ref{fig:dataset}B).

\medskip
\noindent\textbf{Match-ups.  }
We then generated LLM responses to all 30 samples collected using the 5 candidate prompt templates.
Because we used a within-subjects design for our study procedure (Sect.~\ref{sec:study_procedure}), each participant would see half of the match-ups with metrics and half without.
To ensure we collected enough data points in each condition to effectively aggregate task performance and identify decision-making patterns, we choose to have every participant exposed to all of the sample inputs (30) as a match-up, maximizing the number of match-ups per condition (15).
This meant we needed to strategically pair candidate prompt templates to ensure each prompt template showed up the same number of times across both conditions.

For 30 sample inputs and 5 LLM prompts, we pseudo-randomly assigned 2 LLM prompt template candidates to each sample input, ensuring an equal number of candidate pairings across all 30 inputs.
To do this, we generated 12 total LLM responses for each candidate prompt ($30$ inputs $\div$ $5$ prompts $=6$ responses per prompt $\times$ $2$ responses per match-up $=12$ total responses).
These 12 responses were then distributed across 4 candidate pairings per template ($1\choose4$ $=4$) and repeated 3 times ($5\choose2$ $=10$ unique pairings; $30$ inputs $\div$ $10$ unique pairings $=3$ repeated pairings).
For example, the pairings for LLM prompt template candidate 2 would be $2-1$, $2-3$, $2-4$, and $2-5$ (4 pairings).
To evenly distribute all $10$ unique pairings across $30$ inputs, each pairing was assigned to $3$ inputs, resulting in $4\times3=12$ total responses generated for each LLM.
Finally, we ran our metric violation algorithms and recorded violations for each response in all 30 match-ups (Sect.~\ref{sec:metric_algorithms}).

\begin{figure}[!t]
    \centering
    \includegraphics[width=\linewidth]{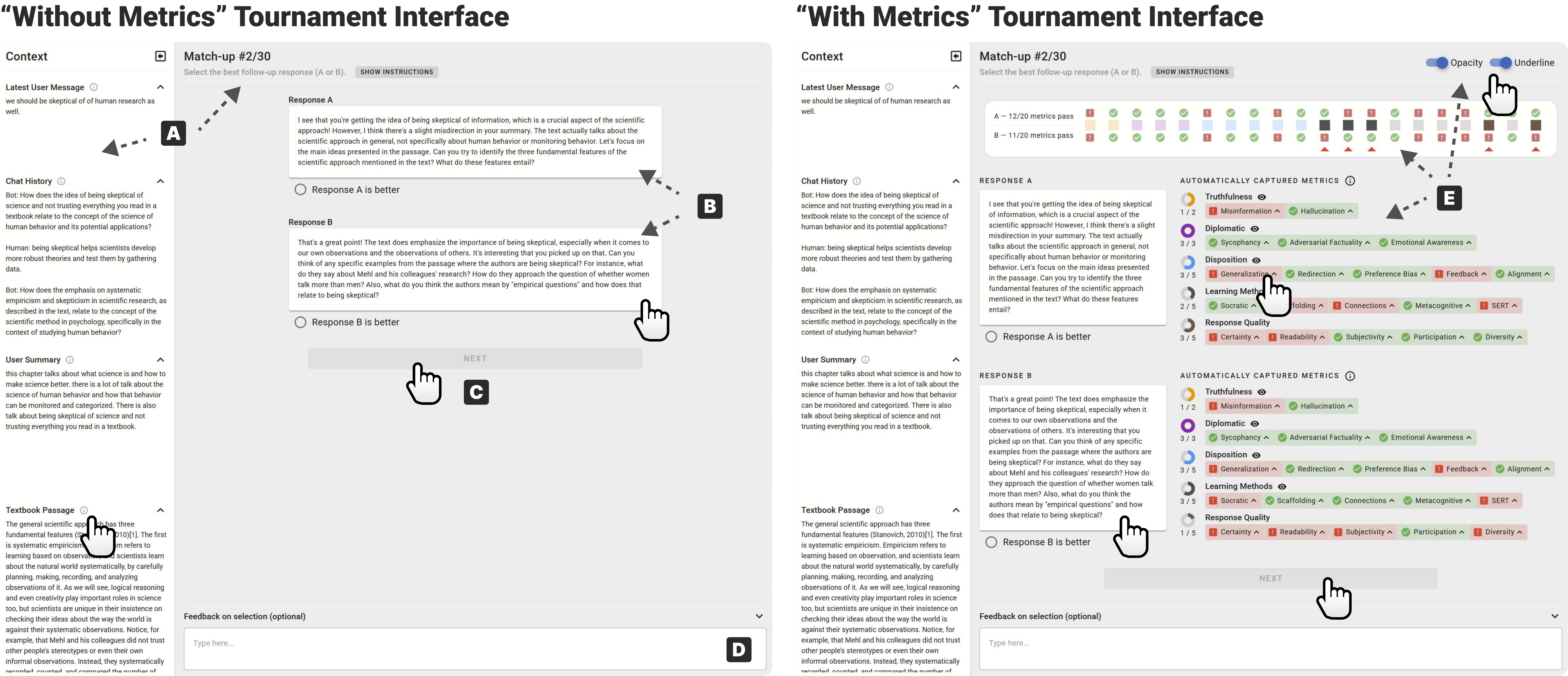}
    \caption{%
        The tournament interfaces we developed as our two study conditions.
        The ``Without Metrics'' interface features a baseline prompt tournament -style interface, showing only the match-up inputs (A) and LLM responses (B).
        Raters click a response and the next button (C) to continue, with the option to provide feedback on their decision (D).
        The ``With Metrics'' interface features the same layout as the baseline, but with the addition of our trustworthiness metrics and visualizations (E).
    }%
    \label{fig:conditions}
    \Description{%
      Accessibility -- TODO
    }%
\end{figure}

\subsubsection{Study Conditions}
\label{sec:study_conditions}
To isolate the effects of using metrics and visualizations on rater decision-making during a prompt tournament, we designed two equivalent interface conditions as shown in Fig.~\ref{fig:conditions} -- one interface without metrics visible, and one interface with metrics visible.

\medskip
\noindent\textbf{Baseline -- ``Without Metrics'' Interface.  }
Our baseline ``Without Metrics'' condition did not feature any metrics or visualizations.
This design mirrors the existing prompt tournament setup that our collaborators currently use, towards ensuring ecological validity \cite{Holmes:2026:PromptTournament}.
The interface is laid out with the match-up inputs on the left (Fig.~\ref{fig:conditions}A), which shows the textbook context, student summary, chat history, and user message from Fig.~\ref{fig:dataset}A.
Raters can click on information icons or the ``show instructions'' button to refer back to the tournament rubric and how each input parameter was collected.
The two LLM responses generated from the match-up inputs on the left are shown in the middle of the screen (Fig.~\ref{fig:conditions}B).
Raters select one of the two LLM responses, then press the ``Next'' button (Fig.~\ref{fig:conditions}C) to continue the tournament.
We also provided a free-text feedback box (Fig.~\ref{fig:conditions}D) for raters to provide commentary on their decision as they work.

\medskip
\noindent\textbf{Experimental -- ``With Metrics'' Interface.  }
Our experimental ``With Metrics'' condition inherits the same interactions and layout as the baseline, but adds one additional element -- the interactive visualizations of the trustworthiness metrics calculated for each LLM response are added to the interface (Fig.~\ref{fig:conditions}E), which we describe in Sect.~\ref{sec:visualizations}.

\subsubsection{Participants}
\label{sec:study_participants}
We recruited $12$ learning engineer collaborators as participants for the tournament (P$1-12$).
All $4$ of the participants in the first two phases of the co-design participated in the tournament; we recruited an additional $8$ collaborators to reach empirical saturation.
All participants had prior experience in developing educational technologies embedded with LLMs, including the digital textbook referenced in this study.
No participants had been exposed to the final visualizations or metrics prior to starting the tournament, to ensure ecological validity.

\subsubsection{Task}
\label{sec:study_task}
\begin{figure}[!t]
    \centering
    \includegraphics[width=\linewidth]{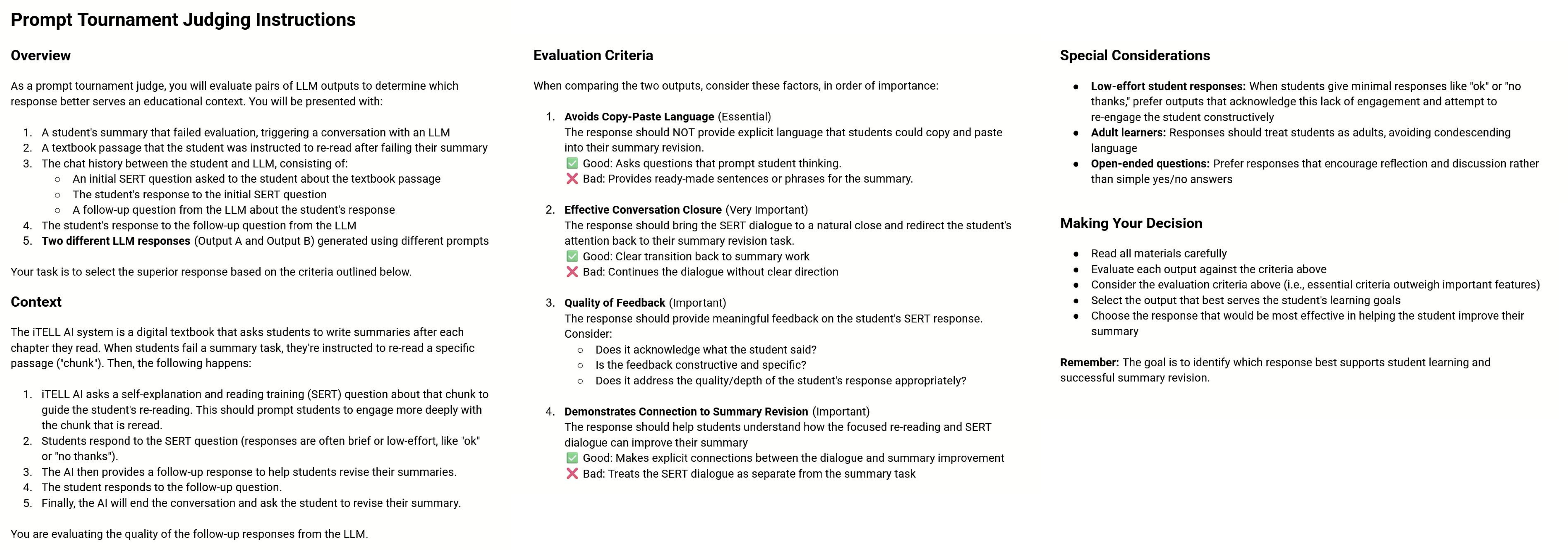}
    \caption{%
        The rubric for the prompt tournament, describing the study task for each rater.
        The rubric was accessible to raters at any time throughout the duration of the study.
    }%
    \label{fig:task}
    \Description{%
      Accessibility -- TODO
    }%
\end{figure}

The central task for each participant was to pick the best LLM response between choice A or choice B. We co-designed a rubric for evaluating match-ups with our learning engineering collaborators, shown in Fig.~\ref{fig:task}.
Raters were not instructed on how to weigh the rubric criteria versus the trustworthiness criteria.
We intentionally allowed for potential conflict between rubric (pedagogical) requirements and trustworthiness measures, as it is unclear what role trustworthiness will play in helping learning engineers decide on the ``best'' responses.
Our results (Sect.~\ref{sec:tournament}) demonstrate this conflict when measuring inter-rater reliability and through participants think-aloud feedback, where we characterized insights into how trustworthiness measures help reveal tensions in evaluating potential pedagogical disruptions across multiple expert perspectives.

The rubric criteria were primarily informed by recurring issues observed in LLM-generated responses during prior textbook deployments.
The structured dialogue sequence centers on self-explanation reading training (SERT) \cite{McNamara:2004:SERT}, and the rubric criteria reflect pedagogical goals appropriate to that context.
The rubric establishes four criteria in descending order of importance. 
First, and most essential, responses must avoid producing copy-paste language: explicit phrases or sentences that students could directly insert into their summary revision without engaging in their own thinking.
This criterion reflects the system's core commitment to student-driven revision, where the LLM's role is to scaffold reflection rather than perform the cognitive work for the student.
Second, responses must effectively close the conversation and redirect students back to their summary revision task, since the response being evaluated is the final turn before students return to the textbook interface.
Third, responses should provide meaningful, constructive feedback on the student's most recent message, acknowledging what the student said, addressing the quality and depth of their engagement, and offering specific rather than generic guidance.
Fourth, responses should demonstrate a connection between the preceding dialogue and the summary revision task, helping students understand how the conversation can inform their revision.

The rubric also includes special considerations for handling low-effort student responses (e.g., ``ok'' or ``no thanks''), which were common in the collected data, and for maintaining an appropriate tone for adult learners.
Raters were instructed to weigh criteria by their stated priority when the two responses differed along multiple dimensions.

\subsubsection{Procedure}
\label{sec:study_procedure}
We split our study into two stages.
First, participants asynchronously rated all 30 match-ups using our tournament interfaces on their own devices, without a study moderator present.
After, they joined a synchronous 30-minute video conferencing call with a study moderator for a semi-structured interview.
We conducted the prompt tournament asynchronously to increase ecological validity -- our learning engineer collaborators regularly host tournaments asynchronously, where raters are free to start the tournament on their own time.
However, to ensure comparable study times across participants, we asked that each participant finish their rating tasks within 1 hour for this study.
Each participant was compensated \$$20$ USD via an Amazon gift card.

\begin{figure}[!t]
    \centering
    \includegraphics[width=\linewidth]{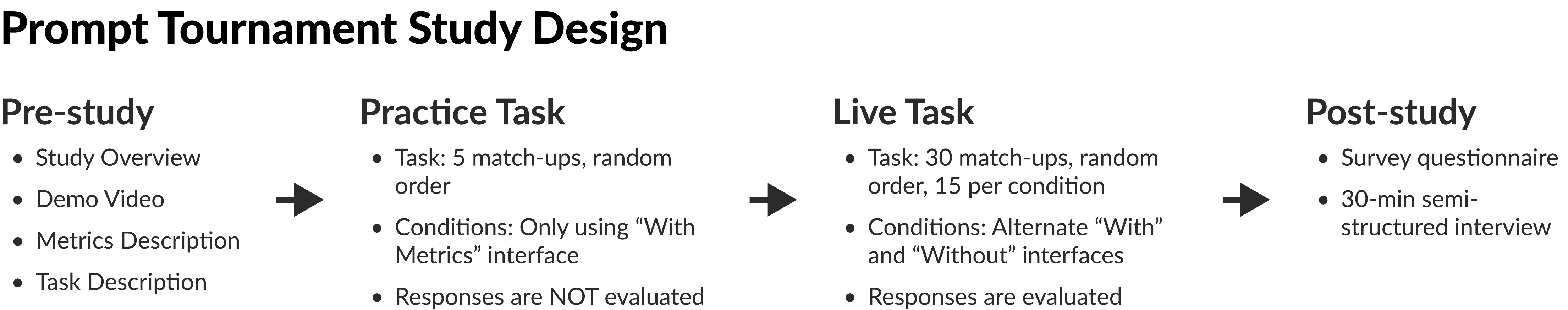}
    \caption{%
        An overview of our study design for conducting an LLM prompt evaluation tournament.
    }%
    \label{fig:study_flow}
    \Description{%
      Accessibility -- TODO
    }%
\end{figure}

An overview of our overall study procedure is shown in Fig.~\ref{fig:study_flow}.
During the first stage, participants first watched a brief video demonstration of the visualization interface features, as well as read an overview of the new metrics and tasks they would perform.
Then, participants completed several practice rating tasks using the ``With Metrics'' interface to get familiar with the new features and overcome learning effects.
These tasks are not included in the results. 
After practice, each participant completed all 30 rating tasks.
For each rating task, participants were shown either the ``Without Metrics'' or ``With Metrics'' interface in an alternating order.
After finishing their ratings, participants then completed a post-study usability survey, capturing the frequency, usability, and usefulness of each metric and visualization.
Finally, participants reached out to the study staff to schedule a 30-minute semi-structured debrief interview within a week of completing the rating tasks.
During the interview, a study moderator discussed with the participant their process, insights, and feedback from the tournament, collecting qualitative quotes that were later analyzed.

We followed a within-subjects study design, in which each participant was exposed to both the ``Without Metrics'' interface and the ``With Metrics'' interface during the live task.
This ensured we could make comparable judgments on task performance between conditions, as well as elicit participant feedback on the benefits and drawbacks of using the metrics and visualizations compared to a baseline experience.
To combat decision fatigue and rushing, we alternated each interface condition after every match-up (i.e., participants would always see the ``With Metrics'' interface after using the ``Without Metrics'' interface, and vice versa).
To reduce ordering effects, we randomly assigned which interface condition would be shown for the first match-up, the order of all match-ups per participant, and which LLM response would be shown as choice A or B per match-up.

\subsubsection{Analysis Plan}
\label{sec:study_analysis}
We conducted a mixed-method analysis using qualitative responses from the post-study interview and in-situ feedback panels, as well as quantitative post-study survey results.

\begin{itemize}
    \item Following Dragicevic \cite{Dragicevic:2016:HCIStats}, we used bootstrapped $95\%$ confidence intervals with $10,000$ resamples. For a given confidence level and sample size, CI width increases with increasing variability; results are considered significant if CIs do not overlap. We report all means and CIs in the text as follows: ($mean$, $[CI]$). We also visualize the raw data and $95$\% CI in several figures as point plots with error bars over top of the raw data as transparent dots.
    \item We performed inter-rater reliability (IRR) analysis \cite{Diaz:2023:IRR} to measure the degree of agreement among participants when selecting LLM responses. Each task presented a rotating pair of prompt options rather than a fixed set of categories, violating the assumptions of several traditional IRR metrics. Given multiple participants ($n=12$) and the need to treat all participant contributions equally, we selected Krippendorff's alpha, harnessing its robust and flexible features that support nominal data, accommodate multiple raters, and handle structurally unavailable categories without bias.
    \item We conducted an inductive thematic analysis \cite{Boyatzis:1998:ThematicAnalysis} of the participant post-study interview transcripts, using open-coding \cite{Saldana:2009:Coding} to identify emergent themes that were then discussed amongst all authors. We then integrated these codes into our inductive themes, enabling us to cross-reference our qualitative insights against our quantitative results to gain a more holistic and comprehensive interpretation of participant behavior. 
\end{itemize}

%% file: sections/5_metrics.tex
In this section, we describe the results of our co-design process with learning engineers to curate a set of metrics for evaluating the trustworthiness of LLMs used in education.

\subsection{Domain Expert Feedback}
\label{sec:feedback_metrics}
Starting from our initial set of curated measures of LLM trustworthiness (Sect.~\ref{sec:phase1_initial}), we first conducted several whiteboarding-style co-design sessions with our learning engineer collaborators over the course of a month (Sect.~\ref{sec:phase1_interviews}) to iteratively refine and align on a useful set of metrics.
We distilled the engineer's feedback by removing, adjusting, and synthesizing measures into new categories (i.e., ``metrics'') that group measures by their pedagogical objectives.

\medskip
\noindent\textbf{Removing unnecessary measures.  }
Several trust-based measures were considered irrelevant, not useful, and or difficult to operationalize, including those bundled under \textbf{Privacy}, \textbf{Safety}, \textbf{Robustness}, and \textbf{Implicit/Explicit Ethics}.
Both privacy and robustness are more important factors in terms of model selection than prompt engineering.
It is difficult to fix issues such as resilience to noise and leakage without changing the underlying model or context of deployment.
Further, safety was not a paramount concern, as the user population of interest for our collaborators is adult learners.
There is less emphasis on making sure the LLM is not toxic and cannot be jailbroken, mirroring the education-specific challenges we uncovered.

\medskip
\noindent\textbf{Adjusting existing measures.  }
Some of the trust-based measures needed adjustment.
\textbf{Truthfulness} was considered important; however the challenge in an educational context is determining what is considered a hallucination or misinformation.
For example, consider an LLM engineered to engage the user in active reading and to encourage seeking ideas.
If the LLM asks for context outside of the classroom materials, it could be flagged as misinformation or hallucination when they may actually be relevant and useful to the active reading process.
One idea was to measure truthfulness in terms of relevance to the learning objective, rather than the source material.
This would help learners know whether the exchange is useful for their learning outcomes or has been derailed. 
In terms of \textbf{Fairness}, our collaborators took inspiration from measuring stereotypes and disparagement.
Because safety and ethics were not as much of a concern, they adapted these measures into new ones.
Measuring stereotypes was adapted into generalization, where the user and LLM may both be on-topic but the LLM is not responding to what the user said, potentially over-generalizing and taking the user off-track.
Measuring disparagement was adapted into redirection, where the LLM changes the topic when the user is still on-topic.

\medskip
\noindent\textbf{Suggesting missing measures.  }
Some new trust-based measures were suggested that were missing from our initial list.
The suggested measures were intentionally lower-inference compared to the larger concepts of ``Truthfulness'' and ``Fairness'' that are hard to measure one-shot with classic NLP approaches.
\textbf{Feedback} would be useful for measuring how encouraging or praising the model is, helping to clarify the disposition of the model.
For example, this measure could capture the inclusion of phrases like ``good job'' or ``try harder''. \textbf{Alignment} would provide a measure of how much the model matches what the student is talking about.
Is the LLM addressing the user or simply reiterating the context?
This measure was suggested to replace the task-based \textbf{Relevance} as a new trust-based measure.
Finally, a new task-based measure was suggested.
\textbf{Readability} was considered important in a pedagogical context, as the LLM may not ``get to the point'' quickly enough or include too many complex terms, bloating the response with jargon.

\medskip
\noindent\textbf{Synthesizing combinations of measures.  }
Based on the removed, adjusted and new measures, several categories of measures were synthesized.
In terms of trust-based measures, \textbf{Diplomacy} was suggested as a combination of truthfulness and ethics measures including sycophancy, adversarial factuality, and emotional awareness.
The response should appropriately question the user without taking them off the rails or agreeing with the learner when it should not.
Our collaborators felt that measuring how much the LLM adapts to the user could be a more relevant way to operationalize certain aspects of truthfulness and ethics towards measuring trustworthiness in education.
Similarly, \textbf{Disposition} was suggested to measure fairness requirements.
The LLM should remain impartial to the user while encouraging them to make progress.
Disposition measures would assess how the LLM engages the user, whether it is helpful, positive, and encouraging of the user staying on topic, without overgeneralizing or exhibiting preference bias.
This category could also include the new measures of feedback and alignment proposed above.
\textbf{Response Quality} was considered as a combination of the task-based measures including topical relevance, coherence, and response structure.
This metric could also incorporate readability.
Finally, the learning-based measures were considered equally important but may not all be relevant to every LLM response.
In other words, the learning outcomes of one type of response may not translate to another.
This suggests a general category of \textbf{Learning Methods} that could be facilitated by evaluating the adherence of the response to the learning objectives.
Our collaborators suggested that an LLM-as-a-Judge \cite{Zheng:2023:LLMJudge} could be a viable measurement tactic.

\subsection{Organizing Measures Into Metrics}
\label{sec:revised_metrics}
\begin{figure}[!t]
    \centering
    \includegraphics[width=\linewidth]{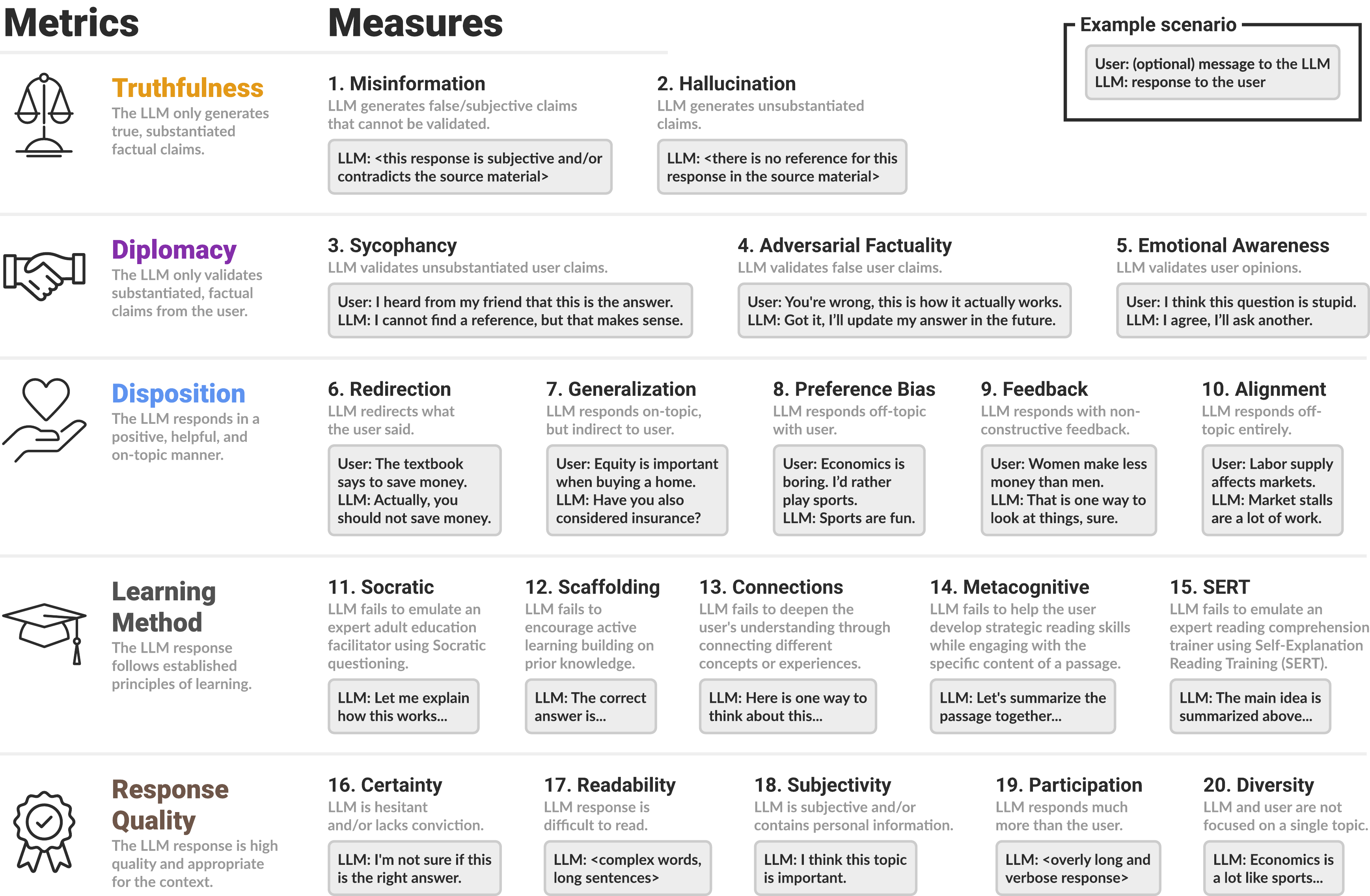}
    \caption{%
        An overview of our final 5 trustworthiness metrics, composed of 20 individual measures.
        Our metrics capture different, complementary aspects of the learning process when students are interacting with an LLM agent in an intelligent digital textbook.
    }%
    \label{fig:metrics_final}
    \Description{%
      Accessibility -- TODO
    }%
\end{figure}

We then distilled the insights from our collaborators to organize our initial measures into new categories of LLM trustworthiness evaluation.
These new categories then became the final set of 5 trustworthiness metrics we contribute, consisting of 20 individual measures (Fig.~\ref{fig:metrics_final}).
Our final set of metrics capture different, complementary aspects of the learning process when students are interacting with LLMs in a classroom setting.
By organizing the measures into higher-level metrics, our framework offers new perspectives on how to define and evaluate LLM trustworthiness.
Future work can easily expand on or adjust our initial categories and measures to adapt to specific educational contexts in which AI are interacting with learners.

\begin{enumerate}
    \item \textbf{Truthfulness. } The LLM should only generate true, substantiated, and factual claims (C4, G2). This metric includes measures of misinformation and hallucination, adjusted to focus on the learning objective rather than a ground-truth answer. This metric considers whether the LLM is responding with information that is relevant to what the student needs to learn. For example, is it giving the student misinformation about what is relevant for the test, or hallucinating important topics that are not in the chapter?
    \item \textbf{Diplomacy. } When responding, the LLM should only validate substantiated, factual claims from the user (C3, C4, G2, G3). Diplomacy combines measures of sycophancy, adversarial factuality, and emotional awareness to measure whether the LLM response attempts to keep the learner engaged in making appropriate responses without discouraging them or adjusting its response to cater to the student.
    \item \textbf{Disposition. } The LLM should respond in a positive, helpful, and on-topic manner (C3, G3). Disposition measures several aspects of fairness including generalization, redirection, and preference bias, as well as constructive feedback and alignment with task. This ensures the LLM engages the user in a helpful, positive, and encouraging manner fitting for educational purposes.
    \item \textbf{Learning Methods. } The LLM responses should follow established principles of learning (C2, C3, G3). This metric captures dimensions of various learning methods that the LLM should adhere to, such as following the Socratic method or engaging the Metacognitive learning process. It is intentionally left open-ended to accommodate changing requirements of learning objectives.
    \item \textbf{Response Quality. } The LLM response should be high-quality and appropriate for the context (C1, C2, G1). Complementing the trust-based and learning-based objectives of the prompts, the response quality also plays an important pedagogical role. Responses should be coherent, relevant, structurally sound, and not too dense with information as to overwhelm the learner. This keeps learners focused and engaged on the task without derailing them.
\end{enumerate}

\begin{figure}[!t]
    \centering
    \includegraphics[width=\linewidth]{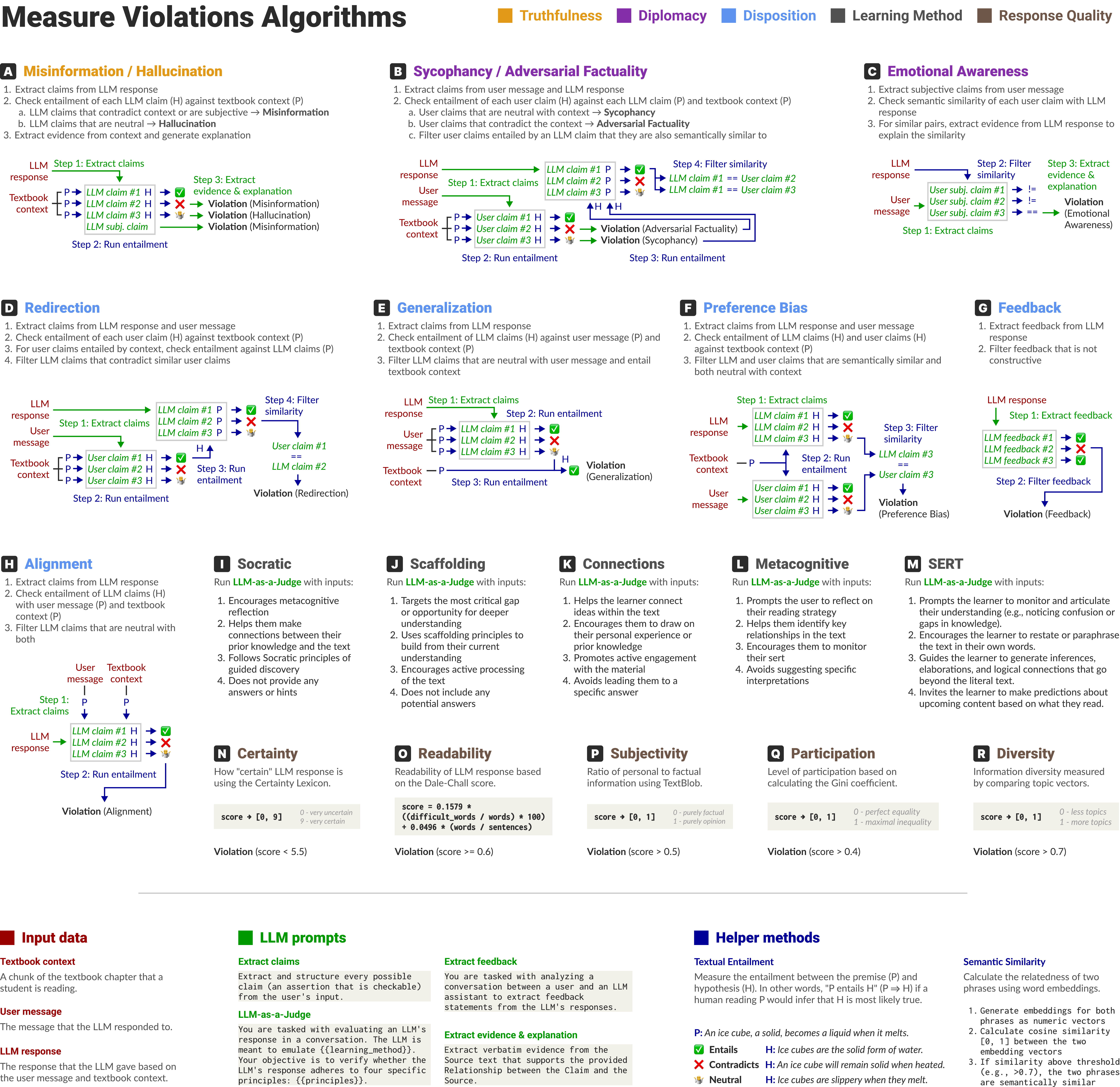}
    \caption{%
        A breakdown of how we calculate violations of all 20 measures.
    }%
    \label{fig:metrics_algorithms}
    \Description{%
      Accessibility -- TODO
    }%
\end{figure}

\begin{figure}[!t]
    \centering
    \includegraphics[width=\linewidth]{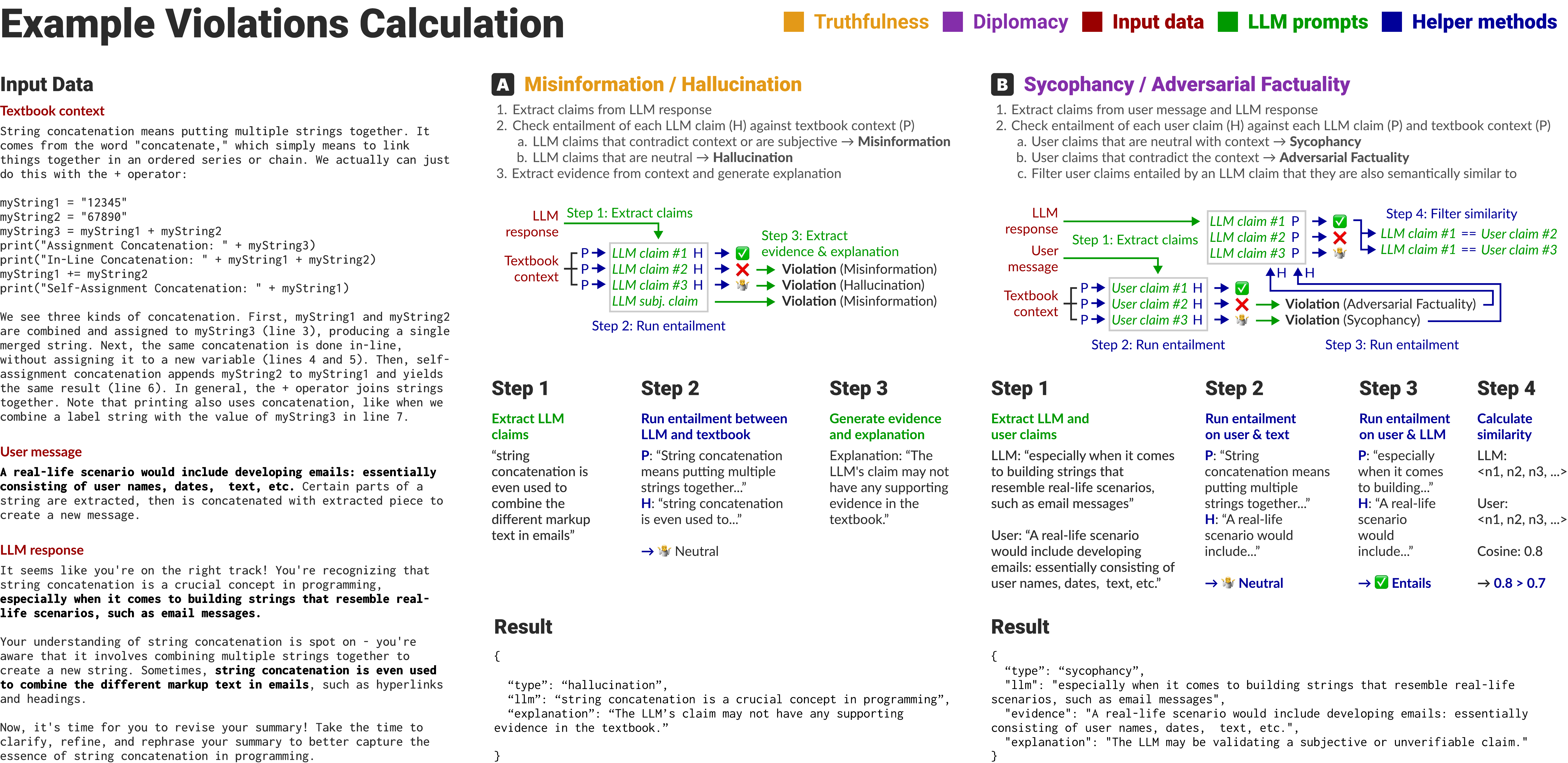}
    \caption{%
        An example of calculating violations of Hallucination and Sycophancy measures on a sample input.
    }%
    \label{fig:metrics_example}
    \Description{%
      Accessibility -- TODO
    }%
\end{figure}

\subsection{Measuring the Metrics}
\label{sec:metric_algorithms}
Our metrics are designed to capture the higher-level pedagogical objectives of measuring LLM trustworthiness in education.
To implement these metrics in practice, we adapted existing measures and, where none existed, created new measures that return quantitative measures.
Some measures from existing metric categories were adapted from Huang et al. \cite{Huang:2024:TrustLLM} to work for an educational context.
The new metrics required entirely new measures to be developed.
Fig.~\ref{fig:metrics_algorithms} gives a detailed breakdown of all 20 measures and how they are calculated.
We show an example of calculating Hallucination and Sycophancy measures on a sample input in Fig.~\ref{fig:metrics_example}.
Our metrics and measures are designed to align with the context of this study, namely evaluating LLM agent responses to learner-sourced conversations from an intelligent digital textbook (Sect.~\ref{sec:background}).

Across all measures, our learning engineer collaborators requested that we measure the \textbf{violation} of each metric specifically.
This means that the measures were developed to identify when a violation of a concept occurs.
During prompt tournaments, one of the central tasks is to identify failure cases as evidence for selecting the best LLM responses.
As our collaborators regularly participate in tournaments to evaluate their LLM prompts, we aimed to increase ecological validity by designing the measures for this use case.
However, each of these measures could easily be adapted to identify the converse -- when an LLM response is satisfying a particular concept.

Finally, implementing our measures to generate discrete versus continuous values was debated.
Our collaborators eventually agreed that in an initial evaluation like this, \textbf{binary} measures would be more approachable to gauge their usefulness in their workflows.
For example, there were questions about how measures such as hallucinations, fairness, and relevance could be measured on a scale.
Rather, it made more sense to think of them as discrete binary indicators, and if there were issues with a given response, to indicate that as a violation.

\medskip
\noindent\textbf{Extracting and comparing claims.  }
Most of our measures work by extracting the claims an LLM response makes and comparing them against the learner's context.
We define a claim as an assertion that can be checked, following work on automated fact-checking \cite{Guo:2022:FactChecking}, and we include this definition in our prompts verbatim.
The pipeline performs the following operations in order.

\begin{enumerate}
    \item \textbf{Extract.} The pipeline takes an LLM response or a learner message as input and extracts each claim it makes, flagging any claim that is subjective and therefore cannot be checked.
    \item \textbf{Compare.} The pipeline pairs each claim with a source text and runs textual entailment, which labels the source as entailing the claim, contradicting it, or being neutral toward it. We compare every claim against four sources: the textbook passage, the learner's summary, the chat history, and the learner's latest message. Comparing against these sources rather than a general standard of correctness is what grounds our measures in the learning context (G2). Where a measure depends on two claims being about the same topic, the pipeline also compares their embeddings for semantic similarity.
    \item \textbf{Explain.} For each claim a measure flags, the pipeline takes the claim, its source, and the relationship between them as input, then extracts the span of the source that justifies that relationship along with a short explanation. These spans and explanations are what our visualizations render inline (G5).
\end{enumerate}

Neutral entailment is central to several of our measures.
A source that is neutral toward a claim neither supports nor refutes it, which is how an unsupported claim surfaces.
For example, when the textbook passage is neutral toward a claim in an LLM response, the response has introduced material the learner was never asked to read.

\medskip
\noindent\textbf{Measuring Truthfulness and Diplomacy.  }
Truthfulness compares the LLM's own claims against the textbook passage (Fig.~\ref{fig:metrics_algorithms}A).
\textbf{Misinformation} is violated when a claim contradicts the passage or is subjective, and \textbf{Hallucination} is violated when the passage is neutral toward the claim.
For example, an LLM may tell a learner that a case study proves cause and effect when the chapter states it cannot, violating Misinformation, or introduce a statistical test the chapter never covers, violating Hallucination.
\textbf{Sycophancy} and \textbf{Adversarial Factuality} are both violated when the LLM affirms a learner claim that the passage does not support, and differ only in whether the passage is neutral toward that claim or contradicts it (Fig.~\ref{fig:metrics_algorithms}B).
For example, a learner may assert that a study proved causation when the chapter describes only a correlation, and an LLM that agrees has validated a misconception the learner will carry into their revision.
\textbf{Emotional Awareness} is violated when the LLM takes up an opinion the learner expressed (Fig.~\ref{fig:metrics_algorithms}C).
Our collaborators asked for this measure because acknowledging how a learner feels can either sustain engagement or divert the conversation, and they wanted to see when it happened.

\medskip
\noindent\textbf{Measuring Disposition.  }
Disposition measures whether the LLM stayed with the learner (G3).
\textbf{Redirection} is violated when the LLM contradicts a learner who was on topic (Fig.~\ref{fig:metrics_algorithms}D), and \textbf{Generalization} is violated when a response is supported by the passage but is neutral toward what the learner said (Fig.~\ref{fig:metrics_algorithms}E).
\textbf{Preference Bias} is violated when the passage is neutral toward both the LLM and the learner, indicating the two have moved off topic together (Fig.~\ref{fig:metrics_algorithms}F).
Our collaborators proposed two additional measures of Disposition.
\textbf{Feedback} extracts the compliments, critiques, and advice in a response and is violated when any of them is not specific and actionable (Fig.~\ref{fig:metrics_algorithms}G).
\textbf{Alignment} is violated when a claim is neutral toward all four sources at once, indicating it relates to nothing in the conversation, the summary, or the passage (Fig.~\ref{fig:metrics_algorithms}H).
For example, an LLM that closes by recommending a study technique never mentioned in the chapter or the dialogue violates Alignment, even though the advice may be sound.

\medskip
\noindent\textbf{Measuring Learning Methods.  }
The five learning methods resist claim-level comparison, so we measure them by prompting a generative LLM to act as a judge \cite{Zheng:2023:LLMJudge}.
For each method, we wrote four principles describing what a response following that method should do.
The judge takes the passage, the dialogue, and the response as input, then returns any principle the response violates along with the text responsible.
We measure \textbf{Socratic} (Fig.~\ref{fig:metrics_algorithms}I), \textbf{Scaffolding} (Fig.~\ref{fig:metrics_algorithms}J), \textbf{Connections} (Fig.~\ref{fig:metrics_algorithms}K), \textbf{Metacognitive} (Fig.~\ref{fig:metrics_algorithms}L), and \textbf{SERT} (Fig.~\ref{fig:metrics_algorithms}M) this way.
For example, the Socratic judge checks that a response encourages reflection and guided discovery, and returns a violation when the response supplies an answer instead.
Writing the principles out let our collaborators inspect and revise the criteria directly, which mattered because what counts as scaffolding is a pedagogical judgment rather than a property of the text.

\medskip
\noindent\textbf{Measuring Response Quality.  }
The five response quality measures come from the Team Communication Toolkit, a library of validated conversational features \cite{Hu:2025:TeamCommTools}, and each is violated when its feature crosses a threshold.
\textbf{Certainty} is violated by a hedging, noncommittal response (Fig.~\ref{fig:metrics_algorithms}N), and \textbf{Readability} is violated when a response scores as difficult on the Dale-Chall formula \cite{Chall:1995:DaleChall}, which our collaborators associated with jargon and bloat (Fig.~\ref{fig:metrics_algorithms}O).
\textbf{Subjectivity} is violated by a response weighted toward opinion over fact (Fig.~\ref{fig:metrics_algorithms}P).
The remaining two measure the conversation rather than the response alone.
\textbf{Participation} is violated when one party does most of the talking (Fig.~\ref{fig:metrics_algorithms}Q), and \textbf{Diversity} is violated when the conversation has spread across many topics (Fig.~\ref{fig:metrics_algorithms}R).
For example, an LLM that answers at length while the learner replies in single words registers as uneven participation, which our collaborators read as a sign the learner has disengaged.
However, we set the thresholds for these last two measures from the distribution of our own dataset rather than an external standard, so they require recalibration before being applied to another corpus.

\medskip
\noindent\textbf{Implementation.  }
We run claim extraction, feedback extraction, evidence extraction, and all five learning method judges by prompting a generative LLM (OpenAI's \texttt{gpt-5} in our implementation), and we compute semantic similarity from \texttt{text-embedding-3-large} embeddings.
For textual entailment we use a DeBERTa-v3-large model \cite{He:2021:DeBERTa} fine-tuned on several natural language inference corpora, an approach that transfers well to new classification tasks with little task-specific data \cite{Laurer:2024:BERTNLI}.
We accept an entailment label only above $50\%$ confidence, and we treat two texts as semantically similar above a cosine similarity of $0.5$.
Our prompts are included in the supplemental material.

%% file: sections/6_visualizations.tex
In this section, we describe the results of our co-design process with learning engineers to design visualizations that can summarize violations of LLM trustworthiness metrics and measures.

\subsection{Domain Expert Feedback}
\label{sec:visualization_feedback}
Starting from our initial visualization designs (Sect.~\ref{sec:phase2_initial}), we conducted several whiteboarding-style co-design sessions with our learning engineer collaborators over the course of a month (Sect.~\ref{sec:phase2_interviews}).
We showed our learning engineer collaborators low-fidelity sketches of visualization prototypes to get design feedback and align the metric feedback with the visualization design.
Several themes emerged in the context of LLM evaluation in education.

\medskip
\noindent\textbf{Visual overviews of metrics. }
Based on domain expert feedback collected when developing the metrics, we realized that the metrics would need to be \textbf{binned} by category, and that would create some interesting visualization challenges.
For example, participants felt the metrics should be visually distinct within bins.
There was tension between when to \textbf{aggregate} individual measures and when to show more information at once.
Our collaborators eventually settled on showing a high-level overview of the bins without details, essentially a binary issue indicator.
This view would highlight any issues that needed addressing in a particular bin, and allow them to expand that bin on-demand.
Comparing between bins was not as interesting, such as in a table view.
Relatedly, many participants felt the \textbf{weighting} of the sub-components was unequal and that the visuals should correlate with the importance of the measure.
In fact, many felt a table view would confuse or conflate the weighting of the measures.
Participants liked the views that \textbf{explained} how each response either addressed or violated a metric.
For example, the Learning Methods metrics use LLM-as-a-Judge \cite{Zheng:2023:LLMJudge} to generate natural language examples that exemplify where the response is failing a particular metric, and why that is.
This helped participants calibrate the pedagogical alignment of the LLM with the metric and know which phrases may be associated with certain metrics, building trust in the system.
Overall, our collaborators wanted metrics being operationalized visually more as signals for whether to think about that dimension of the response, rather than as a guide to understand how to evaluate the response.

\medskip
\noindent\textbf{Visual comparison of metrics across LLM responses. }
Comparison features were considered generally useful to our collaborators.
However, instead of just comparing metrics between responses, participants also wanted to compare metrics against the entire dataset holistically.
For example, basic statistics such as number of violations compared with the dataset average were useful to highlight and compare pairwise.
In general, comparison of metrics within a single response (e.g., comparing violations of Truthfulness to Diplomacy) was not a particularly useful task, as each metric was considered independent of the other.
In other words, knowing how much more ``truthful'' the response was than ``diplomatic'' was irrelevant -- only whether a response had violated being ``truthful'' or ``diplomatic'' at all, and whether this required an explanation.

\medskip
\noindent\textbf{Visual highlighting inline in the response text. }
Text highlighting was considered one of the most useful features to our collaborators.
They suggested several complementary uses of text highlighting.
First, highlighting the position of building blocks of a response in the text.
For example, are questions up-front or in the back? Are main ideas mixed with questions, or explained first? What is the objective of the first sentence? Where are the affirmations located?
Pedagogically, it is important to align the structure of the response with the learning objectives.
Second, how does the position of the phrases of interest compare between responses?
Different objectives may lend themselves to different structures being valuable.
Dovetailing with the suggested comparison visualizations, can the user compare how many objectives the LLM is trying to accomplish in one response?
Third, aligning the user and LLM responses.
Does the LLM repeat what the user said, or paraphrase? Does the LLM match the cadence, effort-level, or tone of the user, or keep to its own method of responding? Does the LLM acknowledge the user, and where does this occur? In the beginning, at the end?
These requests all highlight the versatility and importance of text highlighting as a key visualization technique for enabling LLM trustworthiness evaluation.

\begin{figure}[!t]
    \centering
    \includegraphics[width=\linewidth]{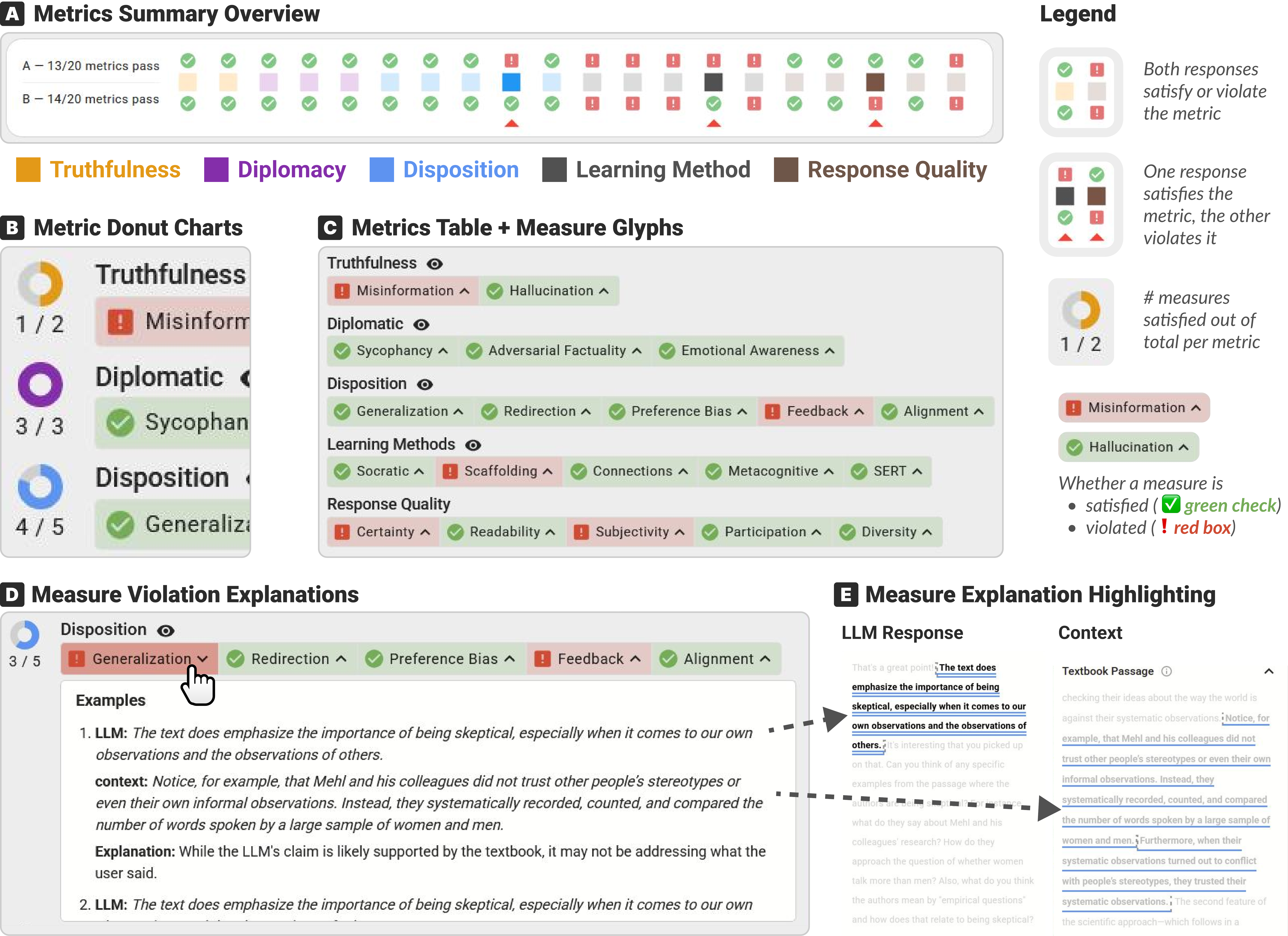}
    \caption{%
        An overview of our final metrics visualizations.
    }%
    \label{fig:vis_final_1}
    \Description{%
      Accessibility -- TODO
    }%
\end{figure}

\begin{figure}[!t]
    \centering
    \includegraphics[width=\linewidth]{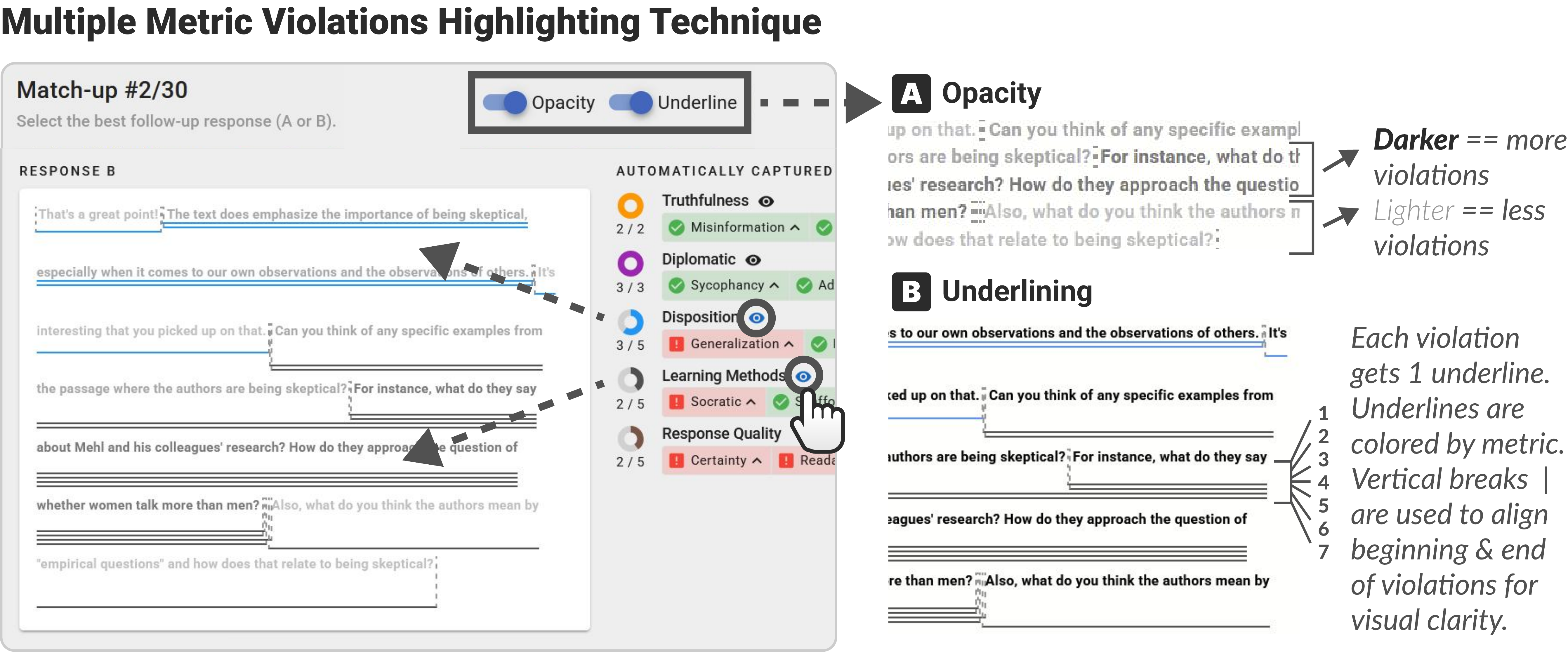}
    \caption{%
        To visualize potentially overlapping measure violations in an LLM response across multiple metrics at the same time, we designed new text highlighting techniques.
        For a given text span labeled as a violation, two different visual encodings can be applied.
        In \textbf{(A)}, text opacity is mapped to the number of times the text span is violated, where darker opacity means more violations.
        In \textbf{(B)}, the text span is underlined for each violation it appears in.
        Underlines are grouped and colored by metric.
        Across both options, a vertical dashed line break is drawn at the start and end of each text span for visual clarity.
    }%
    \label{fig:vis_final_2}
    \Description{%
      Accessibility -- TODO
    }%
\end{figure}

\subsection{Visualizing Trustworthiness Metrics}
\label{sec:revised_visualizations}
Based on the domain expert feedback we collected, we then categorized and applied the insights to revise our initial visualization designs.
Our updated visualization designs are presented in Fig.~\ref{fig:vis_final_1}.
The visualizations each solve unique design challenges in enabling metric-based evaluation of LLM trustworthiness in education, and together they achieve our visualization goals: summarizing violations for comparison (\textbf{G4}) and tracing each violation back to the source text (\textbf{G5}).

\begin{enumerate}
    \item \textbf{Metrics summary overview (G4). } We designed a summary visualization to compare metrics between two LLM responses (Fig.~\ref{fig:vis_final_1}A). The visualization lays out both LLM responses vertically and arranges the measures of each metric horizontally. We use glyphs to indicate whether each response is satisfying or violating a measure. Where responses differ, we include a marker to make the difference more visually salient. These views help users get a broader sense of how the issues relate to the other response and the dataset at large.
    \item \textbf{Metric donut charts (G4). } To summarize how often a single LLM response violates a topical metric, we designed donut charts to show how many measures are violated per metric (Fig.~\ref{fig:vis_final_1}B). These donut charts are shown next to the table of measures to give an overview of metric performance at a glance. We map a unique color scheme for each metric category that is then shared across all of the visualizations. This helps users quickly identify and compare violations across metric categories.
    \item \textbf{Metrics table and measure glyphs (G4). }  We organized measures in a table view horizontally, grouped by metric (Fig.~\ref{fig:vis_final_1}C). Each measure is represented in the table by a binary glyph that indicates whether a response has violated that measure. A table view allow users to quickly scan and identify the kind and amount of violations in each response.
    \item \textbf{Measure violation explanations (G5). } Certain measures generate explanations for why the measure was violated and extract text examples from the context, user message, and LLM response as evidence. We designed a detailed explanations and examples panel that can opened on-demand below any measure in the table (Fig.~\ref{fig:vis_final_1}D). The explanations reveal how a particular metric was measured, how it was violated, and what clauses in the response were responsible for the assessment. These views help users drill down for details on how and why a response might be failing a specific measure without overwhelming them with details up-front.
    \item \textbf{Inline violation highlighting (G5). } To more clearly highlight where in the text data input a measure was violated, we designed a text highlighting technique that applies automatically when opening an explanations panel in the metrics table (Fig.~\ref{fig:vis_final_1}E).
\end{enumerate}

When highlighting explanations and examples of specific measures being violated, there may be more than one violation captured per measure.
Further, our collaborators expressed an interest in comparing the positionality of parts of LLM responses that relate to the measures, such as where the LLM is violating Truthfulness in relation to Disposition.
However, we realized there were no clear guidelines on how to visualize overlapping categorical text spans inline while maintaining the semantic structure of the original text block.
For example, if a part of an LLM response violates both Truthfulness and Disposition, how should that be drawn? How do we encode the number and type of violations to overlapping text spans? How do we ensure visual clarity?
To solve this, we designed two new text highlighting visualization techniques that extend traceability to overlapping violations (\textbf{G5}), shown in Fig.~\ref{fig:vis_final_2}.
We engaged two salient visual encoding channels as dual encodings.

First, we mapped text opacity to the number of violations that a text span appears in (Fig.~\ref{fig:vis_final_2}A).
Darker text spans indicate that phrase has more violations, while lighter text spans indicate less violations.
This encoding scheme aims to associate the trustworthiness of a text span to the opacity -- darker spans have more violations, and therefore are easier to identify and count at a glance.
This also allows users to gauge the positionality of ``untrustworthy'' phrases in an LLM response, such as in the front of the response or at the end, which our collaborators specifically requested.

Second, we mapped text underlines to each individual violation of a text span (Fig.~\ref{fig:vis_final_2}B).
Each violation gets a single underline, colored by associated metric using the color scheme from the metric donut charts.
Underlines are then drawn stacked vertically below text spans, with each line reserved for a single violation.
This technique gives users a more fine-grained view of how many and which measures a text span violates, and allows for visual comparison of the positionality of violations by metric.
For example, the underlines make it easier to see when LLMs continually violate Disposition measures (in blue) at the start of their responses, compared with violating Learning Methods measures (in grey) at the end of their responses.
This distinction can help learning engineers spot specific behaviors in LLM generation to address.

Finally, to improve the visual clarity of both text highlighting techniques, we draw dashed vertical line breaks at the start and end of each text span.

%% file: sections/7_tournament.tex
In this section, we describe the results of our LLM prompt evaluation tournament with $12$ learning engineers comparing $5$ different LLM prompt templates used to generate responses to $30$ real learner conversations.
We ground our analysis in four specific categories of investigation:

\begin{itemize}[topsep=1em, leftmargin=5em]
    \item[\textbf{Sect.~\ref{sec:results_winner}}] Which LLM prompt won? Which LLM prompt was most trustworthy? Did participants agree with each other?
    \item[\textbf{Sect.~\ref{sec:results_factors}}] How did time spent, overall number of metric violations, and individual measure preferences influence decision-making?
    \item[\textbf{Sect.~\ref{sec:results_themes}}] When did participants look at the metrics and visualizations? How did participants use the metrics and visualizations to support their decision-making workflows?
    \item[\textbf{Sect.~\ref{sec:results_usability}}] Which interface features did participants find most/least useful for decision-making?
\end{itemize}

Achieving the desired educational outcomes from selecting the ``best'' prompt template is a complex process spanning multiple criteria, and there is no clear procedure for weighting trustworthiness with respect to rubric criteria.
By introducing trustworthiness metrics and visualizations, we aimed to study the impact of providing a more structured lens for LLM evaluation, towards bridging gaps in conflicting objectives across multiple expert perspectives (Sect.~\ref{sec:methodology_phase1}).
We specifically investigated how the metrics and visualizations influenced expert decision-making behaviors to better understand what LLM behaviors are desirable, identifying areas of disagreement and highlighting important challenges of LLM response evaluation in education.
We expect that our metrics and visualization will help raters to align on selecting the ``best'' LLM responses that also mitigates issues of trustworthiness.

Trustworthiness in this paper is a framework for structuring LLM evaluation and identifying potential pedagogical risks that could undermine trust for downstream users (in this case, learners reading the digital textbook).
Our trustworthiness metrics do not represent the learning engineers' or the learners' underlying trust in the LLM itself.
However, the learning engineers' trust in the metric itself played a decisive role in their decision-making process, which we expand on in Sect.~\ref{sec:results_factors}.
We also acknowledge that individual differences in participant's personalities, level of domain expertise, and lived experiences can influence their tendency to trust and thus affect our ability to discern trusting behavior \cite{Evans:2008:TrustMeasurement}.
This is further compounded by the relationship between reliability and trustworthiness, where it is well-established that the option participants select most frequently in a survey is not always the ``optimal'' choice.
In this study, the prompt templates were all written by experts familiar with the textbook context for LLM deployment, potentially influencing rater's subjective preferences for LLM responses generated by one prompt over another.
Further, any improvements to agreement that the trustworthiness metrics and visualizations foster may come at the expense of other rubric criteria or even aspects of trustworthiness that are not covered by the current visualizations and metrics.
We explore the trade-offs between the participant expectations, the rubric, and trustworthiness factors in Sect.~\ref{sec:results_themes}.

\begin{figure}[!t]
    \centering
    \includegraphics[width=\linewidth]{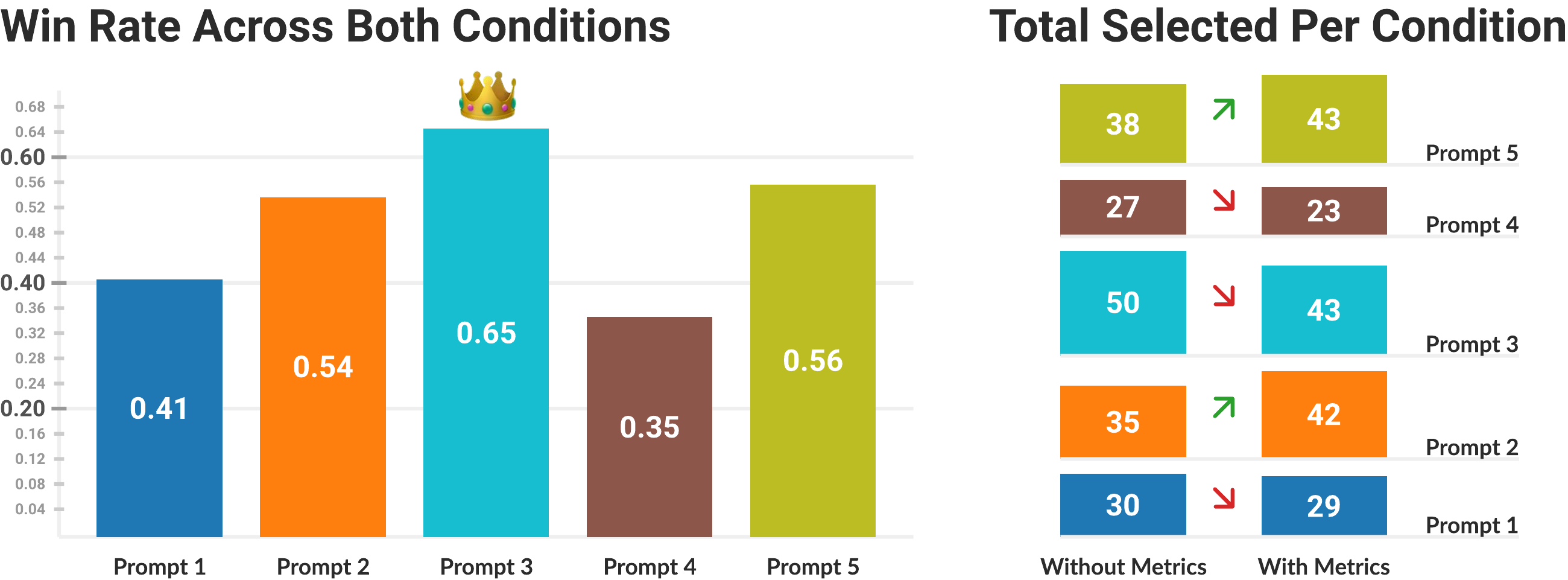}
    \caption{%
        Raw win rates for each LLM prompt template across both conditions (left) and broken down by whether metrics were visible in the interface (right).
        Overall, Prompt 3 was chosen most often and was the clear winner when the metrics were not visible.
        However, when the metrics were visible, Prompt 2 and Prompt 5 were each chosen more often, rivaling Prompt 3 for top spot.
    }%
    \label{fig:win_rates}
    \Description{%
      Accessibility -- TODO
    }%
\end{figure} 

\subsection{Deciding The Winner}
\label{sec:results_winner}
LLM prompt template 3 was the overall winner of the tournament across both interface conditions.
However, Prompt 3 did not produce the most trustworthy responses overall, with the honor going to Prompt 5.
Looking at the effect of the interface conditions reveals a promising pattern -- that when metrics were visible, participants more often chose responses generated with Prompt 5.
This could indicate that metrics and visualizations were not enough to overcome individual preferences, but they did help learning engineers align more often on responses which were deemed more trustworthy.
This is partially confirmed by inter-rater reliability (IRR) analysis.
Overall agreement between conditions was poor to moderate ($\alpha = 0.4344$), but showed improvement when isolating each condition ($\alpha_0 = 0.3987$ without metrics visible; $\alpha_1 = 0.4931$ with metrics visible).

\begin{figure}[!t]
    \centering
    \includegraphics[width=\linewidth]{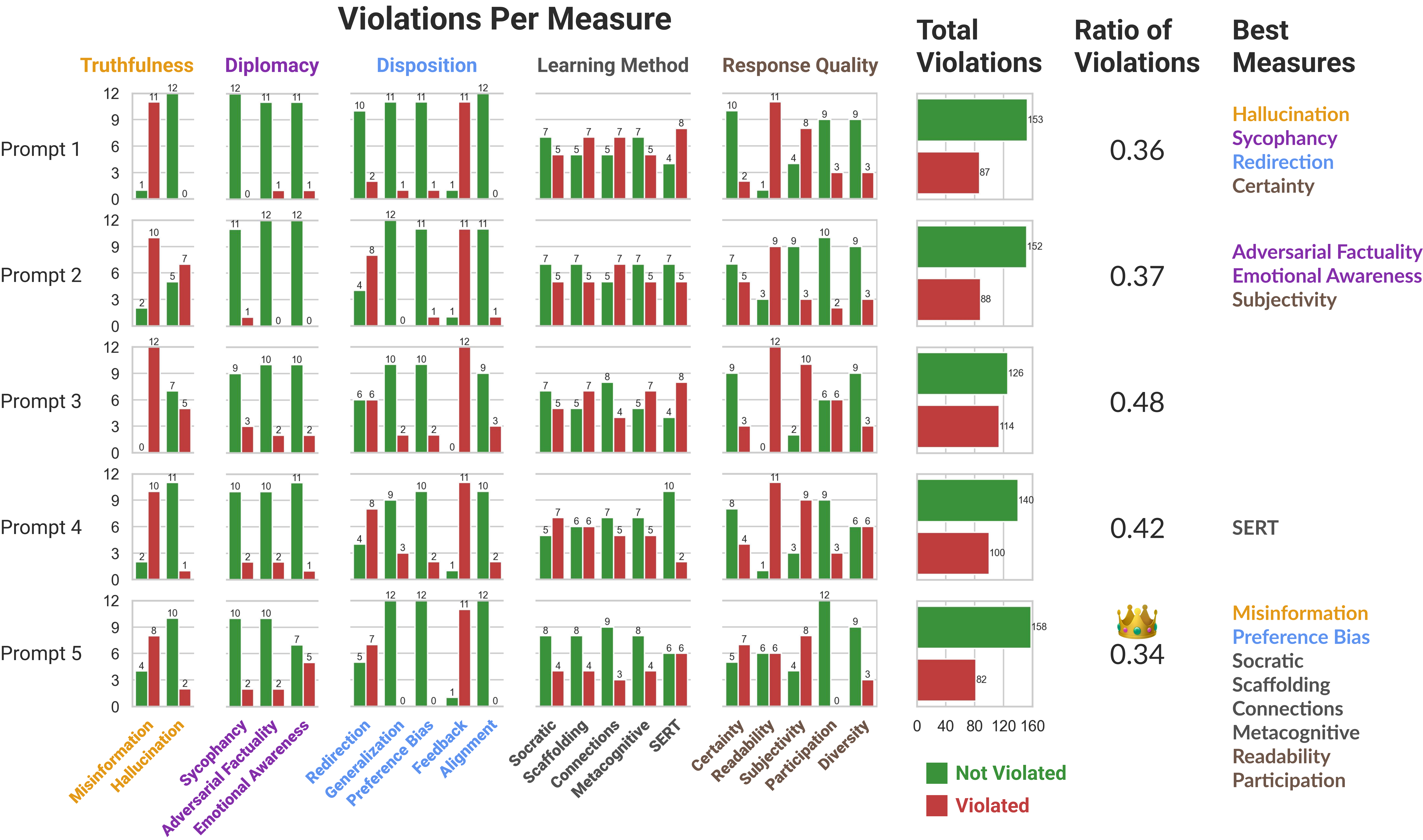}
    \caption{%
        Count of measure violations in all responses per LLM.
        Prompt 3, the winner of the prompt tournament, had the most violations and did not outperform any other prompt template on any measure.
        However, Prompt 2 and Prompt 5 were both selected more often when metrics were visible and had less violations.
        This could indicate that different decision-making factors are competing to select the ``best'' LLM prompt template.
    }%
    \label{fig:violation_total}
    \Description{%
      Accessibility -- TODO
    }%
\end{figure}

\medskip
\noindent\textbf{Prompt 3 won, but only when metrics were not visible. }
Fig.~\ref{fig:win_rates} shows the win rates of each LLM prompt template across both interface conditions as well as within each.
We plotted raw win rates (chosen/total), finding that Prompt 3 was the top performer across both conditions, achieving the highest win-rate at $0.65$.
Conversely, Prompt 4 was the least chosen option, with a win rate of only $0.35$.
Breaking down the selections by interface condition shows a more nuanced picture of the overall ``winner''. 
When the metrics were not visible, Prompt 3 won with clear majority of 50 selections, with Prompt 5 in second position with 38 picks.
Yet when metrics were visible, the selections were more evenly distributed -- Prompt 3 and Prompt 5 are tied with 43 selections each, followed closely by Prompt 2 with 42 selections.
Across both interfaces, Prompt 4 was selected least frequently, indicating overall poor fit for the downstream task.
Next, we explore whether this change in decision-making behavior is related to how trustworthy the LLM responses were.

\medskip
\noindent\textbf{Prompt 5 was the most trustworthy, and was selected more often when metrics were visible. }
Is the winning LLM prompt template the ``most trustworthy'', according to our metrics?
We calculated the number of measure violations (True/False) per metric across all candidate prompt templates.
We then aggregated the total number of measure violations per template and calculated the ratio of violations over total evaluations performed to determine the ``most trustworthy'' prompt template quantitatively.
The results are shown in Fig.~\ref{fig:violation_total}.

Prompt 5 had the least number of total metric violations, making it the ``most'' trustworthy LLM prompt template according to our metrics.
However, comparing against the winning LLMs, we see that Prompt 5 came in second, whereas our ``least'' trustworthy LLM prompt template, Prompt 3, won the tournament.
Further examining each measure across all prompt templates, we identified which measures each LLM prompt template outperformed the others on.
Prompt 5 had the least violations in the most number of measures, particularly succeeding at not violating $4/5$ learning methods (Socratic, Scaffolding, Connections, and Metacognitive).
Prompt 3, on the other hand, did not outperform any other prompt template in any measure.

While the most chosen LLM prompt template was not the most trustworthy, one striking observation is the performance of Prompt 5.
It came in second place in the overall rankings, and when broken down by interface, was selected more often when metrics were visible than when they were not.
Similarly, Prompt 2 had a lower violation ratio than other options and showed an increase in the number of selections when the metrics were visible.
This pattern seems to indicate a dual perspective -- the trustworthiness metrics could be affecting participants decisions, but not enough to sway the majority.
We break down and examine the effects of decision factors that could be at play in the next section.

\medskip
\noindent\textbf{Participant agreement increased when metrics were visible. }
We expected that metrics and visualizations would increase the agreement between participants on which option is best, which we saw partial evidence of for Prompt 5 and Prompt 2.
To measure agreement, we computed inter-rater reliability (IRR) using Krippendorff's alpha comparing participants' LLM selections across all tasks.
The resulting agreement score was $\alpha = 0.4344$.
This score falls below the standard reliability threshold of $0.67$, indicating poor to moderate agreement among participants beyond chance.
However, computing IRR for each interface condition yielded more promising results; without metrics visible, participants agreed less ($\alpha_0 = 0.3987$), yet when metrics were visible, agreement rose ($\alpha_1 = 0.4931$). 
The increase in level of agreement between conditions may suggest that the metrics and visualizations were useful to resolve internal conflict when the rubric objectives were not clearly met by either responses.
Another possible explanation for this rise in consistency is that participants may have placed varying emphasis on the rubric depending on the presence of specific metrics of interest. 
The addition of visualizations could be influencing this behavior by providing quick reference points for flagged metrics that help participants make faster judgments.
The visualizations likely helped participants reconcile their internal reasoning with the metrics indicated, even when the metrics underperformed.
We discuss this effect when we examine how participants used the metrics and visualizations to make decisions in Sect.~\ref{sec:results_themes}.

\subsection{Decision-Making Factors}
\label{sec:results_factors}
Looking at the results of the tournament, we observed a dual-perspective -- participants most often favored responses from the LLM prompt template that likely satisfied the rubric criteria, but agreement on the ``best'' responses improved when trustworthiness metrics and visualizations were visible, producing a more even distribution of ``best'' responses.
Which decision-making factors were responsible for this observation?
We expected that time spent, the number of metric violations, and preferences for individual measures would likely influence behaviors the most.
Time spent showed the least influence, subverting our expectation that participants would take longer to read, interpret, and utilize the visualizations to pick more trustworthy responses.
The amount of metric violations more often led participants to pick more trustworthy responses, although a few participants would filter or ignore some metrics outright, skewing these results.
Finally, individual preferences for measures played a major role, leading to a greater diversity in selected responses based on their trustworthiness, although this came with a strong caveat that any discrepancies in how a measure was supposed to work would cause the learning engineers to distrust using that measure in future match-ups.

\begin{figure}[!t]
    \centering
    \includegraphics[width=\linewidth]{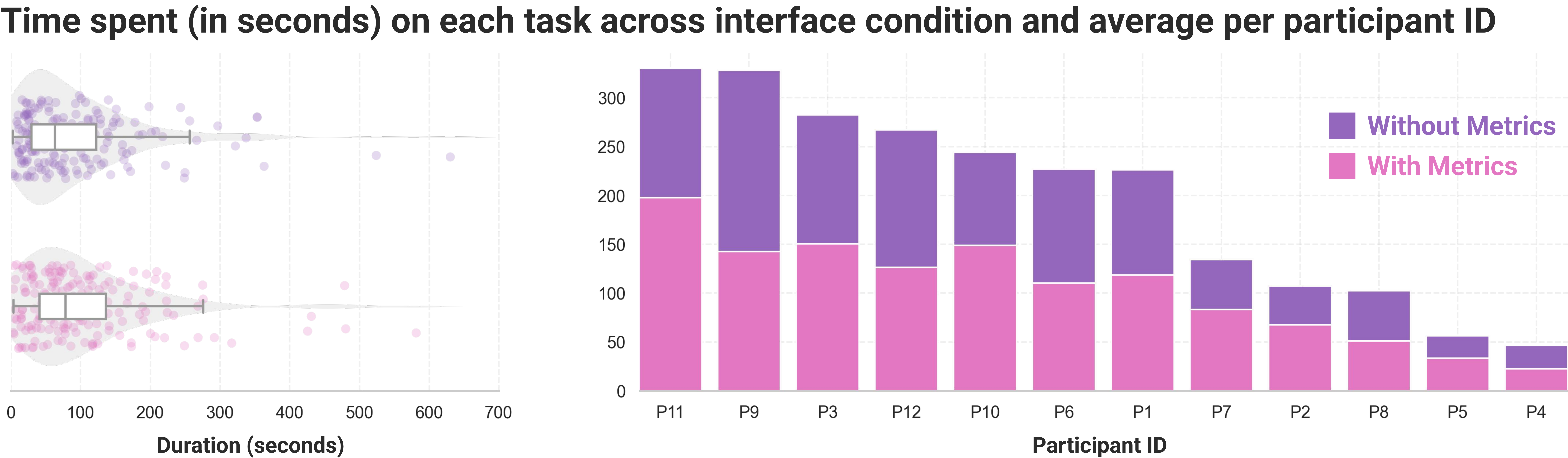}
    \caption{%
        Time taken by the participants to choose a response for each interface condition.
        Overall, having metrics visible did not significantly increase the time it took for participants to select a response.
    }%
    \label{fig:time_spent}
    \Description{%
      Accessibility -- TODO
    }%
\end{figure}

\medskip
\noindent\textbf{Time spent rarely impacted decision-making. }
Fig.~\ref{fig:time_spent} shows the time participants took to choose a response in each interface condition, as well as all times plotted in each condition in a violin plot.
Median time spent per participant ranges from $21$s to $137$s.
Participants spent slightly longer on tasks with metrics visible (median $79.55$s) than without metrics visible (median $62.95$s), and only half of them ($6/12$) were slower when metrics were visible, subverting our expectations that most participants would take much longer to assess the metrics with the visualizations and render a decision.
P$8$, one of the few that took longer, explained that: \textit{``For the first few questions, I would examine metrics in detail, but that was time-consuming. After a while, I spent less time making decisions.''}.
Participant preferences for specific metrics or simply the number of violations are likely more influential than time spent.
Case in point, sometimes the visualizations helped participants make decisions faster.
For P$3$, \textit{``the Donut Chart helped me to quickly understand where I needed to read a response deeper and saved me time overall.''}
This added benefit of visualizing metrics may have influenced the results by encouraging new heuristics to form, such as quickly checking for number of violations as a proxy of trustworthiness.
Alternatively, the task for this tournament may have been too simple, making the need to more carefully evaluate the metrics less important for quickly making decisions.
Next, we investigate if using the amount of violations as a heuristic for decision-making was more influential.

\medskip
\noindent\textbf{Overall metric violations became a fast but unreliable heuristic for decision-making. }
We analyzed whether participants more often chose responses with a higher amount of metric violations (higher-violation) or a lower amount of violations (lower-violation).
We expected that participants would more often choose the LLM option with less total metric violations when using the interface that showed metrics, to minimize total violations overall. 

\begin{figure}[!t]
    \centering
    \includegraphics[width=\linewidth]{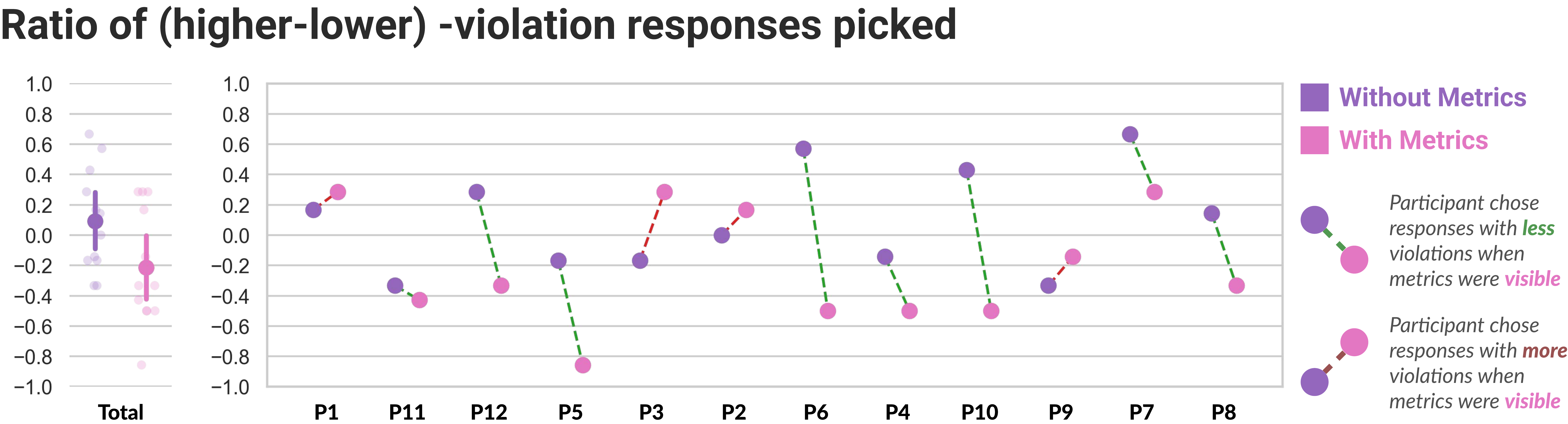}
    \caption{%
        The ratio of (higher - lower) -violation responses chosen over total per participant across interface conditions.
        Higher values indicate higher-violation responses were picked more often, and vice versa.
        Overall, when metrics were visible, participants tended to pick the lower-violation response.
    }%
    \label{fig:violation_ratio}
    \Description{%
      Accessibility -- TODO
    }%
\end{figure}

Fig.~\ref{fig:violation_ratio} illustrates the proportion (part/total) of match-ups where LLM responses with more violations were picked versus less violations were picked.
When metrics were visible, participants more often picked LLM responses with less violations ($-0.214$, [$-0.421$, $-0.002$]) compared to when the metrics were not visible ($0.093$, [$-0.089$, $0.282$]).
We are encouraged to see the results in the ``Without Metrics'' condition, where participants would not know if the response they chose had more or less violations than the other, tended towards a pattern of random choice; i.e., a lower standard deviation centered around a ratio of 0, indicating no preference.
In contrast, the preference when metrics were visible tended away from 0 towards picking lower-violation responses with greater confidence.
Viewing the number of metric violations reported in visualizations like the Donut Chart likely influenced this effect.
For example, P$6$ explained that \textit{``the number of metrics satisfied was a decision factor like any other. I would pick options sometimes that just had more metrics satisfied altogether.``}
As shown in Fig.~\ref{fig:violation_ratio}, P$6$'s ratio shows a strong decreasing trend from selecting higher-violation responses when the metrics weren't visible, to selecting the lower-violation response when the metrics were visible, corroborating this pattern.

At the same time, we sometimes observed cognitive dissonance in participant's expectations regarding number of violations as a decision-making criteria.
Participant P$3$ stated \textit{``my choices were often aligned with the metrics``}, and similarly, P$1$ noted that \textit{``the responses I chose often had more green check marks.``}
However, in Fig.~\ref{fig:violation_ratio} we see both of the participants P$1$ and P$3$ had opposite trends to P$6$, more often choosing higher-violation responses when metrics were visible.
One explanation could be their perception of metric importance -- they may have observed only the metrics \textit{they cared about} with mostly green check marks.
This behavior was more directly expressed by P$2$ and P$9$, who both tended to pick higher-violation responses as well.
Both reported minimal use of metrics for decision-making and a feeling of dissonance between the metrics and rubric.
P$9$ explained that \textit{``most of the metrics were not connected to the judging rubric, so I mostly ignored them.''}
Other participants reported similar tension, such as P$7$: \textit{``There was a tension between following the judging rubric and considering the trustworthiness metrics, especially where they clashed. I would prefer to satisfy the rubric over the trustworthiness metrics.''}.
Others like P$12$ only sometimes found themselves doing this, depending on the response: \textit{``Sometimes I would pick the response that had more metrics violated... if the rest of the response was good, I wouldn't weight the metric violations as much, even if there were more of them.''}.
Filtering or outright ignoring metrics likely also had an impact on the final agreement between participants.

Overall, we see more influence from the number of metric violations as a criteria for decision-making.
Some participants used it as a quick heuristic, while others calculated the number only for the metrics they cared.
Others ignored the number, trusting their own internal judgment.
Next, we investigate the role of personal preferences for specific metrics on the outcomes of decision-making.

\begin{figure}[!t]
    \centering
    \includegraphics[width=\linewidth]{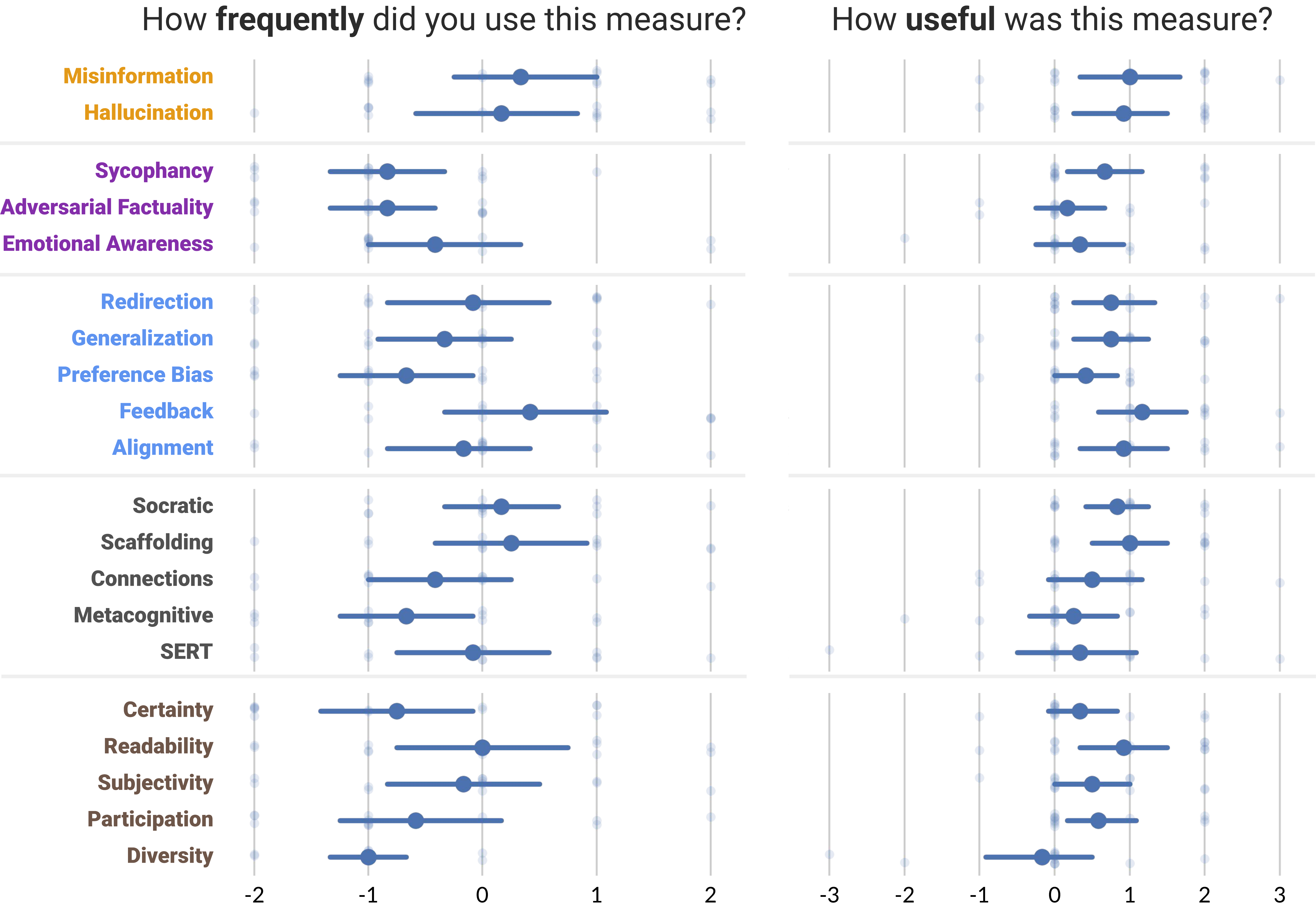}
    \caption{%
        $95\%$ CI around mean self-reported frequency ($[1, 5]$) and usefulness ($[-3, 3]$) of each measure.
        Misinformation, Hallucination, Feedback, Socratic, Scaffolding, and Readability were both frequently used and useful.
        Adversarial Factuality, Preference Bias, Connections, Metacognitive, Certainty, and Diversity were consistently underutilized and considered not as useful.
    }%
    \label{fig:survey_metrics}
    \Description{%
      Accessibility -- TODO
    }%
\end{figure}

\medskip
\noindent\textbf{Individual measure preferences often mirrored decision-making behaviors, but revealed trust issues with some measures. }
We examined self-reported preferences for each measure, as well as how often responses were chosen in spite of each measure being in violation, to better understand how personal preference played a role in decision-making.
We expected that participants would adjust their decision-making to favor choosing LLM responses that did not violate the measures they cared about.

Fig.~\ref{fig:survey_metrics} shows the aggregated self-reported frequency and usefulness scores for each measure.
Misinformation, Hallucination, Feedback, Socratic, Scaffolding, and Readability were both frequently used and useful, whereas Adversarial Factuality, Preference Bias, Connections, Metacognitive, Certainty, and Diversity were consistently underutilized and considered less useful.
Familiarity with the concepts that the measures were evaluating played a major role; e.g., P$4$ explained that \textit{``I didn't pay attention to measures like SERT because I didn't know them well''}, while P$5$ similarly felt they \textit{``didn't have a background in some of the metrics, so I didn't rely on them too much.``}

\begin{figure}[!t]
    \centering
    \includegraphics[width=\linewidth]{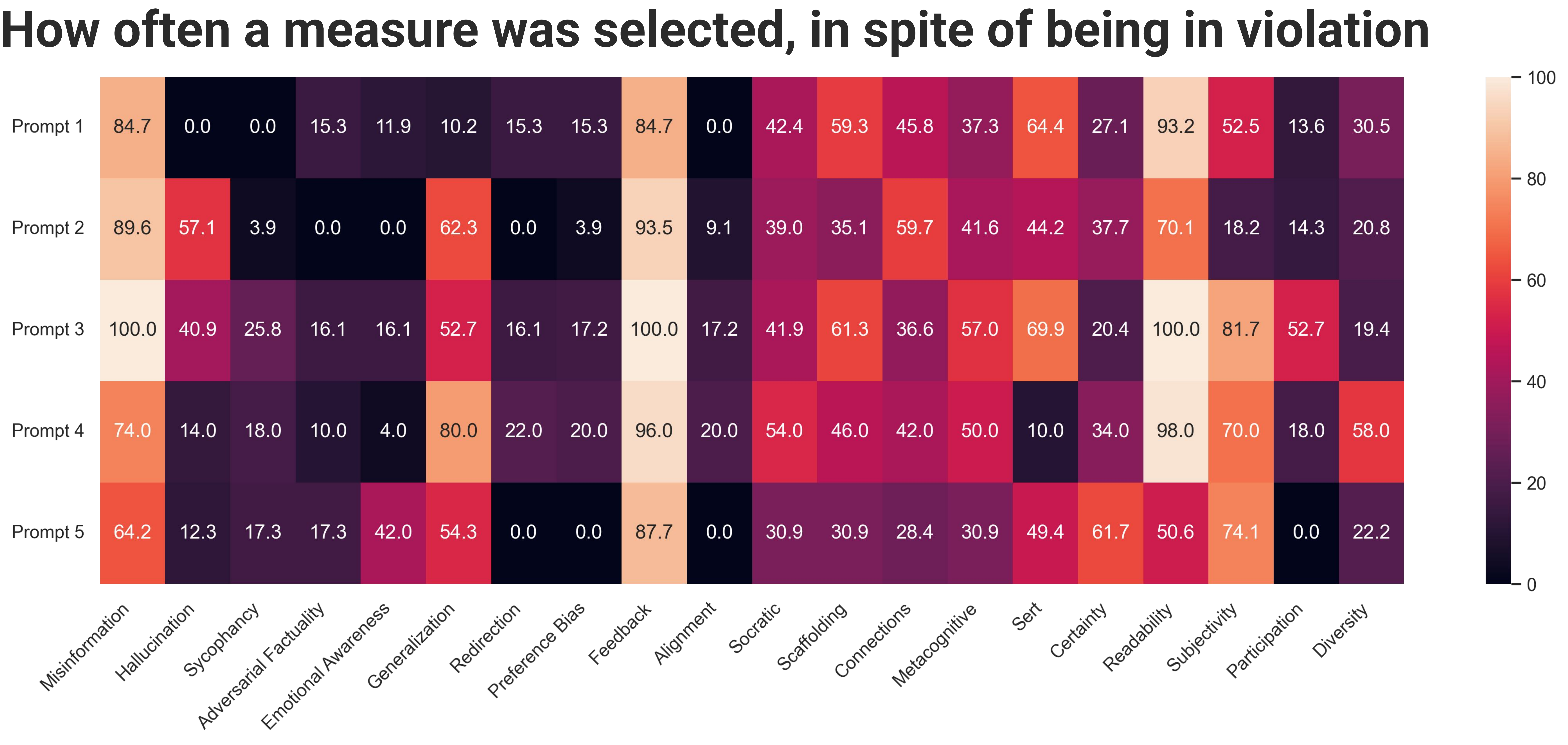}
    \caption{
        The percentage of measure violations in all responses chosen per LLM.
        Higher values and lighter colors indicate that responses were chosen with that measure in violation more often, and vice versa.
        Overall, some responses were consistently chosen with similar amounts of measure violations, potentially indicating these measures were ignored or not important.
        However, some measures (Misinformation, Hallucination, Generalization, Readability, Subjectivity, and Diversity) showed greater variation in whether the response was picked while they were in violation, potentially indicating intentional decision-making at play.
    }
    \label{fig:violation_percentage}
    \Description{%
      Accessibility -- TODO
    }%
\end{figure}

Fig.~\ref{fig:violation_percentage} presents how often responses were selected when each measure was in violation.
In other words, a higher value means the chosen responses violated that measure more often, and vice versa.
These results correspond with our breakdown of measure violations in Fig.~\ref{fig:violation_total}.
We observe certain trends in decision-making that generally correlate with the self-reported preferences for measures in Fig.~\ref{fig:survey_metrics}.
For example, Misinformation, Hallucination, and Readability all show a higher deviation in whether they were in violation when responses were selected, potentially indicating these measures were useful as decision-making criteria.
Other measures show similar deviation, including Emotional Awareness, Generalization, SERT, Certainty, Subjectivity, Participation, and Diversity.
Some measures that were self-reported as less useful, including Adversarial Factuality, Preference Bias, Connections, and Metacognitive, all demonstrated a lower deviation, potentially indicating they may have been more often ignored or were less important for decision-making.

The variation in preference for measures was also influenced by participants' trust in the measure itself.
Some participants were reluctant to use measures they were unfamiliar with, which likely had an influence on the decision-making behavior we observed when examining self-reported usefulness/frequency and the ratio of violations to times selected.
For example, P$1$ reported they \textit{``lost trust in the Socratic measure and some others because I disagreed with how they were measured.''}.
This disagreement was influenced by the quality of the measure; if there were discrepancies in how participants expected the measure to be evaluated and the results it actually returned, cognitive dissonance emerged.
This is particularly important in the field of educational technology, where measures that predict potential pedagogical harms are few and far between.
P$2$ echoed this sentiment, saying \textit{``in our field, it's easy to lose trust if anything is wrong. So, gaining trust is hard when there's a wrong example.''}.

\subsection{Participant Workflows and Behaviors}
\label{sec:results_themes}
We observed variations in decision-making outcomes based on whether the metrics and visualizations were visible.
Importantly, participants did not always chose the LLM response with the best-performing metrics, challenging our expectations.
Thus, we ask: How did participants use the trustworthiness metrics and visualizations to make decisions?
Overall, we find that the human-in-the-loop approach was important for helping participants interpret the metrics in context of the LLM response.
Different participants have their own preference for metrics, but many expressed benefiting from having the metrics as a checklist. 
The visualizations were most useful to overcome the limitations of using metrics alone for decision-making.

\begin{figure}[!t]
    \centering
    \includegraphics[width=\linewidth]{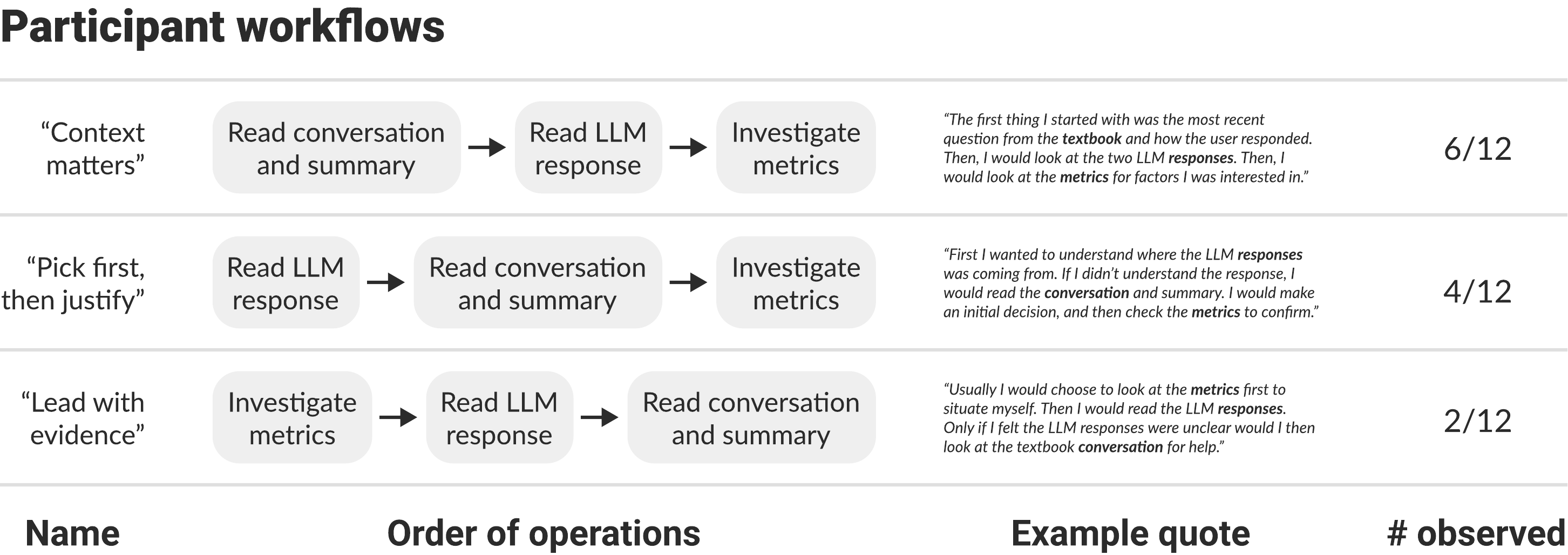}
    \caption{%
        Table of observed participant workflows during the prompt tournament.
    }%
    \label{fig:workflows}
    \Description{%
      Accessibility -- TODO
    }%
\end{figure}

\medskip
\noindent\textbf{Participants often looked at the metrics and visualizations last. }
We conducted an inductive thematic analysis and organized participant quotes related to the operational steps they took to select an LLM response. 
Three distinct workflow patterns emerged, summarized in Fig.~\ref{fig:workflows}.
Overall, most participants would look at the metrics last, only after reading the LLM responses and textbook context, and usually only when they had not already made an initial decision on which response they preferred.
This likely influenced much of utility feedback we got about the metrics and visualizations.

\begin{itemize}
    \item The \textbf{``Context matters''} workflow was the most common workflow we observed in $6/12$ sessions. Participants began by reading the conversation history between the LLM agent and learner in the textbook, then moving on to read the LLM responses. At this point, participants may have already made a decision on which response they preferred; only if they were stuck would they then check the trustworthiness metrics last.
    \item The \textbf{``Pick first, then justify''} workflow was the second most common, observed in $4/12$ sessions. Participants started with reading the LLM responses; then, if the participant could not decide right away, they would review the textbook conversation to clarify any confusion. Similar to ``Context matters'', they would prefer to look at the metrics at the very end, as P$4$ explains: \textit{``I would read/skim responses from A/B, then skim the user message after to see if the prompt was addressing the textbook goal correctly. Only if both were good responses did I look at the metrics.''}.
    \item The \textbf{``Lead with evidence''} workflow, only observed in $2/12$ sessions, featured participants starting with the metrics to ground their evaluation in the data first, before reading the LLM responses and getting additional context. P$7$ explained that \textit{``my impression was the AI was not aligned to what I was thinking, so I wanted to clarify any subjectivity beforehand.''}.
\end{itemize}

\medskip
\noindent\textbf{Metrics resolved conflicting objectives by acting as a procedural checklist. }
We expected that the metrics would foster reflection on which criteria are important for decision-making.
Looking at the outcomes from the tournament, it is evident that each participant has their own internal goals during the tournament.
The metrics helped with managing conflicts in these internal goals and the LLM responses by externalizing the decision-making criteria as a procedural checklist.
 
The metrics played a crucial part in creating a bridge for more consistently applying decision-making heuristics that mattered to each participant. 
P$4$ summarized this benefit, stating: \textit{``the metrics helped me compare along objective dimensions.''}. 
P$1$ felt the metrics drew attention to pedagogical issues that mattered: \textit{``there was a question where I looked at Scaffolding, and the metric pulled out an example that said an LLM response violated Scaffolding. It changed how I look at the responses, and I started looking for Scaffolding more often... the metrics helped me to consider pedagogical concerns.``}.
In this way, the metrics acted like a procedural checklist, helping participants ensure all relevant criteria were considered during judgment.
P$6$ explained that they would \textit{``look at the metrics to find things that I missed initially. I would treat the metrics like a mental checklist, to keep track of things that I need to check''}.
P$8$ describe that \textit{``sometimes I would look at the response first before the metrics. Then I would look at the metrics and realize my internal choice was in conflict with the metrics. Then I would go back and redo my decision process all over again''}. 
We observed the metrics helping participants resolve internal conflicts by foregrounding measures that matter for pedagogical evaluation, towards establishing a more robust and repeatable LLM evaluation process. 

\medskip
\noindent\textbf{Visualizations helped participants reconcile internal reasoning with what the metrics were showing. }
We expected that participants would not blindly follow the metrics, but critically examine any violations for accuracy and consistency, using the visualizations to identify important issues more easily, helping participants make informed decisions.
Every participant noted using one or more of the visualizations, and many participants used several visualizations in different phases of their workflow.
In particular, the visualizations helped participants reconcile their own internal reasoning and conflicts with what the metrics were indicating, even when certain metrics were violation. 

The visualizations helped participants keep track of specific metrics, making to easier to notice violations and address them in their decision-making.
P$7$ explained that, \textit{``the visualizations made it easy to get an overview of the trustworthiness. I could easily determine how much I wanted to rely on the metrics, and it was super easy to ignore them when I wanted to''}.
Comparison between metrics was another common use case of the visualizations.
P$8$ felt that \textit{``the red and green highlighting helped me determine which metrics I had to consider quickly and save time by reading the response to evaluate the response in term of the metric.''}
P$4$ was able to mentally bin categories of evaluation using the metric color scheme: \textit{``The colors really helped separate metrics into groups and how groups passed or failed, helping me make a decision faster.''}
The text underlining also enabled several participants to directly map explanations of metric violations onto LLM responses and the textbook conversation, drawing participants attention to details they may have overlooked initially.
Overall, the visualizations complemented the metrics-based approach by helping participants ground their decision-making process in a more explainable way, highlighting important considerations and ways to resolve internal conflicts.

\begin{figure}[!t]
    \centering
    \includegraphics[width=\linewidth]{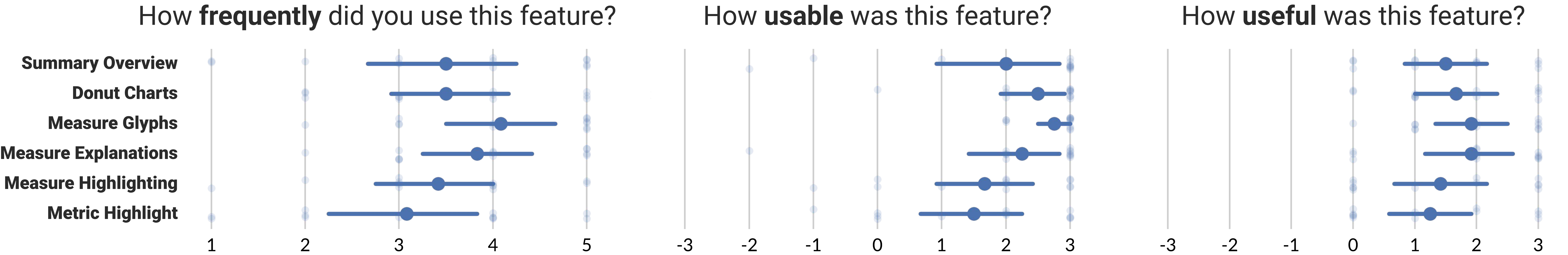}
    \caption{%
        $95\%$ CI around mean self-reported frequency ($[1,5]$), usability ($[-3,3]$), and usefulness ($[-3,3]$) ratings of interface features.
        The glyphs and explanations were used the most often and considered highly usable and useful.
        Both the measure and metric text highlighting were underutilized and had usability issues.
    }%
    \label{fig:usability}
    \Description{%
      Accessibility -- TODO
    }%
\end{figure}

\subsection{Interface Usability}
\label{sec:results_usability}
Which interfaces features did participants find the most or least useful for decision-making?
We asked participants to self-report the frequency, usability, and usefulness of each feature in their post-study survey, described in Fig.~\ref{fig:usability}.
Overall, we find that the metrics summary overview, metrics table and measure glyphs, and metrics donut charts were considered the most useful visualizations, while the measure explanation highlighting was highly divisive.

Representing individual measures as red/green glyphs made it easy for participants to quickly check and assess violations to make their decisions, as P$6$ confirmed: \textit{``the green and red glyphs were very useful; my most used feature!''}
Participants also expressed how they often used the metrics donut charts to guide their decision-making process; e.g., P$3$ felt that \textit{``the donut chart helped me to quickly understand where I needed to read a response deeper and saved me time.''}
P$7$ expressed a similar observation, stating: \textit{``the donut chart provided a unique way to check for consistency and reliability between and across responses. I really liked it.'}.
P$11$ also praised the donut charts, saying that \textit{``the donut charts were helping me compare the dimensions of categories that I cared about between the responses. If the donuts were not agreeing, I wanted to dig in deeper''}.
The metrics summary overview garnered mixed reactions: some participants found it very useful, leveraging it as one of their main tools, whereas others found it confusing to use.
Some used it as a major part of their process, like P$1$: \textit{``The summary chart was helpful to find a pattern in how many metrics were satisfied overall for each response and compare''}.
On the other hand, some completely ignored it and even found it confusing, like P$2$: \textit{``I didn't use the summary vis at all or the highlight all feature''} and P$3$: \textit{``The summary visualization was confusing, I didn't use it as much''}.

The most divisive feature by far was the measure explanation highlighting.
For some, the highlighting made it easier to spot potentially pedagogical issues directly in LLM responses, helping participants make decisions more easily.
For example, P$6$ noted they \textit{``didn't care for the opacity, but I did like the underline feature and the highlight all eye feature.``}
On the other hand, the amount of underlining and text highlighting made it confusing for some participants to read the LLM response.
As P$9$ explains, \textit{``I didn't use the text underlining much, it was confusing. I did look at the explanations and examples a few times when I wanted more context and having the visual of where in the text that example is was very helpful when i used it, but I didn't use it very much. The all highlighting was less usable because it was harder to read.``}.
We consider ways to improve on our preliminary visualization designs in Sect.~\ref{sec:limitations_future_work}.

%% file: sections/8_discussion.tex
By collaborating with learning engineers integrating a pedagogically-aware LLM agent into an intelligent digital textbook, our work identified a core challenge of evaluating LLMs in education: managing conflicting expert perspectives on whether LLM responses satisfy pedagogical objectives (Sect.~\ref{sec:design_process}).
To bridge this gap, we adapted trustworthiness criteria from the ML community as a lens for LLM evaluation in education through a three-stage co-design process (Sect.~\ref{sec:methodology}).
First, we filtered, grouped, and synthesized trustworthiness measures into metrics that measure important pedagogical objectives for LLM responses (Sect.~\ref{sec:metrics}).
Second, we designed visualizations of the trustworthiness metrics that summarize and explain how LLM responses are violating criteria by mapping examples directly onto responses (Sect.~\ref{sec:visualizations}).
Third, we investigated the impact of trustworthiness metrics and visualizations on decision-making through an LLM prompt evaluation tournament, grounding our work in a real evaluation process used by our domain expert collaborators (Sect.~\ref{sec:tournament}).

In Sect.~\ref{sec:does_it_matter}, we discuss the pedagogical benefits we observed in applying a trustworthiness lens to LLM evaluation in education.
Then, in Sect.~\ref{sec:design_guidelines}, we synthesize these insights into design guidelines for future LLM evaluation tools that better foreground pedagogical risks, while providing greater decision-support for domain experts.
Finally, in Sect.~\ref{sec:limitations_future_work}, we discuss the limitations of our approach and suggest future directions for research.

\subsection{How Do Trustworthiness Metrics and Visualizations Help Learning Engineers?}
\label{sec:does_it_matter}
We observed four main benefits of including trustworthiness metrics and visualizations in the LLM evaluation process: (1) increasing expert agreement on what constitutes the ``best'' LLM responses; (2) expanding awareness of potential pedagogical risks that may have been overlooked or underexplored before; (3) helping resolve conflicting pedagogical objectives to streamline decision-making; and (4) enabling participants to reconcile their internal reasoning with problematic LLM behaviors.
Taken together, these results suggest that trustworthiness is not only a promising lens for surfacing potential pedagogical risks, but could also enable practical structuring mechanisms for helping experts streamline decision-making in LLM evaluation.

Our work demonstrated that operationalizing trustworthiness as a set of interpretable metrics has the potential to improve both the efficiency and reliability of expert decision-making.
When these metrics were made visible, expert agreement increased, indicating that metrics can serve as a shared baseline for pedagogical alignment even in the presence of existing rubrics.
Notably, specific metrics including \textbf{Misinformation}, \textbf{Hallucination}, \textbf{Emotional Awareness}, \textbf{Generalization}, \textbf{Certainty}, \textbf{Subjectivity}, and \textbf{Participation} were consistently reported as useful for shaping decisions, while also demonstrating positive effects on overall decision-making outcomes.
This suggests that identifying a core subset of metrics could foster more robust decision support in practice, such as in a hybrid automated and human-in-the-loop evaluation pipeline.
Automated screening could first identify clear trustworthiness violations (e.g., misinformation or hallucinations), reducing the burden of manual review, while flagging more ambiguous or complex violations that learning engineers can resolve using metric-informed visualizations as decision-support tools.
Dividing labor can preserve the efficiency gains we observed while still maintaining sensitivity to nuanced pedagogical concerns that are difficult to fully automate.

At the same time, evaluating LLM trustworthiness remains inherently complex.
Treating trustworthiness metrics as a flat, fully automated optimization objective risks overlooking subtle, ambiguous, and context-dependent issues that are pervasive in educational settings.
In this regard, our trustworthiness visualizations played a crucial role in breaking down this complexity into more interpretable and consumable forms.
Rather than enforcing convergence, these visualizations supported experts in aligning their internal decision-making criteria with broader evaluation objectives.
For example, when participants encountered problematic or conflicting LLM behaviors, the visualizations enabled them to reconcile pedagogical disruptions and reason explicitly about trade-offs between responses.
In other words, by foregrounding dimensions such as accuracy versus emotional awareness, participants were better equipped to make nuanced and defensible judgments about pedagogical impact.
Notably, this process led to a greater diversity in what was considered the ``best'' response when metrics were visible.
We interpret this increased variation not as inconsistency, but as evidence of deeper engagement and more reflective judgment, particularly in cases where pedagogical risks conflicted with rubric-defined objectives.
Beyond immediate decision support, this capability has important downstream implications.
By surfacing which dimensions most strongly influence expert judgments, visualizations can help learning engineers identify and formalize previously underexplored pedagogical criteria.
These insights can then inform future iterations of evaluation workflows, creating a compounding effect in which both the criteria and processes for assessing LLMs become progressively more robust and pedagogically aligned.

Overall, our findings suggest that trustworthiness metrics are most effective not as standalone automation targets, but as components of interactive, human-centered evaluation systems.
By combining automated screening with visualization-supported expert review, future LLM evaluation frameworks can better balance scalability with the nuanced judgment required for pedagogical alignment.
In the next section, we provide practical design guidelines for building such systems.

\subsection{Design Guidelines For Future LLM Evaluation Tools}
\label{sec:design_guidelines}
From our study results, we synthesized design guidelines to enhance future LLM evaluation tools.

\medskip
\noindent\textbf{Focus expert attention through configurable metric views. }
Evaluation interfaces should enable experts to prioritize and tailor trustworthiness metrics based on their domain knowledge and evaluative goals.
Our findings show that participants did not treat all metrics equally; instead, they selectively focused on those they deemed most pedagogically relevant (e.g., Misinformation vs Emotional Awareness).
To support this behavior, systems should provide configurable features such as metric weighting and the ability to toggle metrics on or off.
These controls may help reduce cognitive load and allow experts to align the interface with their internal decision criteria.
At the same time, designs should balance flexibility with accountability, as excessive user control may result in overlooked violations.
Supporting lightweight reflection mechanisms (e.g., prompting users to revisit flagged but ignored metrics) may help mitigate this risk.

\medskip
\noindent\textbf{Integrate metrics and rubrics into a unified evaluation framework. }
Trust in trustworthiness metrics partially depends on their alignment with existing evaluative frameworks such as rubrics.
When perceived inconsistencies arise, experts were unclear about how to reconcile their decision-making.
One way to resolve this conflict could be more tightly integrating metrics with rubric-based evaluation.
Future systems should explicitly map metrics to rubric dimensions, co-visualize them, or provide explanations of how metric scores operationalize rubric criteria, in addition to addressing trustworthiness concerns.
Such integration can reduce friction, improve trust in automated measures, and support more coherent decision-making.

\medskip
\noindent\textbf{Leverage metrics to iteratively refine evaluation criteria. }
Trustworthiness metrics and visualizations can serve not only as evaluation tools, but also as mechanisms for discovering new pedagogical criteria.
By making influential dimensions of decision-making visible, these systems enable learning engineers to identify which factors matter most in practice especially those not captured in existing rubrics.
Designers should support workflows that capture, analyze, and iterate on these insights, enabling evaluation frameworks to evolve over time.
This creates a compounding effect in which both the criteria and processes for assessing LLM outputs become increasingly robust and pedagogically aligned.

\medskip
\noindent\textbf{Prioritize comparative visualizations to encourage trade-off reasoning. }
Evaluation tools should support reflective judgment rather than force agreement.
Visualizations in our study enabled participants to reason through trade-offs (e.g., accuracy versus Emotional Awareness) and reconcile conflicting pedagogical objectives, leading to deeper engagement rather than inconsistency.
To achieve this, not all visual encodings equally supported trustworthiness evaluation.
Participants strongly preferred visualizations that facilitated comparison and synthesis across multiple dimensions such as summary views, glyphs, and aggregated charts (like the metric donut charts), while single-metric encodings like opacity or underlining produced mixed reactions.
This suggests that effective designs should prioritize visualizations that make trade-offs and cross-metric relationships explicit.
At the same time, designers should carefully manage how overlapping metrics are represented, as multiple highlighting techniques can introduce visual clutter or ambiguity.
Balancing clarity and expressiveness in multi-dimensional visualizations remains a key design challenge in supporting pluralistic, defensible decision-making.

\subsection{Limitations and Future Work}
\label{sec:limitations_future_work}

We evaluated our trustworthiness metrics and visualizations using an LLM prompt evaluation tournament \cite{Holmes:2026:PromptTournament} to maintain ecological validity with our learning engineer collaborators' current practices.
However, participants often struggled to balance objectives between the tournament rubric, a familiar decision-support tool, with the unfamiliar metrics and visualizations.
Future work should investigate how repeated use of trustworthiness criteria impacts trust, reliability, and consistency in decision-making behavior, as well as expanding support for broader range of objective criteria for LLM evaluation, including more open-ended pedagogical goals.
For example, consider using trustworthiness criteria to evaluate LLM prompts that guide open-ended exploration of a dataset in a data science class.
Additionally, we only used LLaMa3 \cite{Touvron:2023:LLaMa} to further maintain ecological validity with the current textbook framework.
Future studies should explore other LLMs, especially those fine-tuned for educational contexts such as Gemini \cite{Imran:2024:GeminiEducation}.

We adapted existing trustworthiness measures from the ML literature to avoid confounds from creating new metrics, but their accuracy applied to educational contexts remains untested.
Evaluating these measures on prior datasets, such as prior tournament datasets, could clarify the preferences and distrust among participants we observed in our study.
Moreover, despite the benefits that trustworthiness metrics and visualization introduced, participants identified some aspects that were less effective and warrant further consideration.
Managing several metrics added additional cognitive overhead for some participants, either drawing their attention away from the task to consider the trade-offs of metrics, or causing them to start ignoring metrics in future tasks.
Similarly, while all participants benefited from at least one form of visual encoding, they noted diminishing returns as the number of encodings increased.
Future work should investigate the cognitive load demands of introducing metrics and visualization -based decision support into LLM evaluation.

During co-design, our collaborators suggested additional visual analytics approaches for expanded decision support that did not make the final cut.
They likely would have introduced greater training overhead before starting the tournament, potentially impacting ecological validity.
Expanding on a visual analytics approach for evaluating LLMs in education remains an open challenge.
For example, temporal views could track participants' answers in the tournament over time to help them understand patterns related to their choices of responses, particularly related to under- or over-relying on specific metrics.
This builds on related work in guidance for visual analytics \cite{Ceneda:2017:GuidanceinVA}.
Future tools could further synthesize insights from these responses and visualize their relatedness, shared text, and distribution of metrics.
This type of analysis has been explored before in the visualization community for LLM evaluation; e.g., in \cite{Kahng:2024:LLMComparator}.

\begin{figure}[!t]
    \centering
    \includegraphics[width=\linewidth]{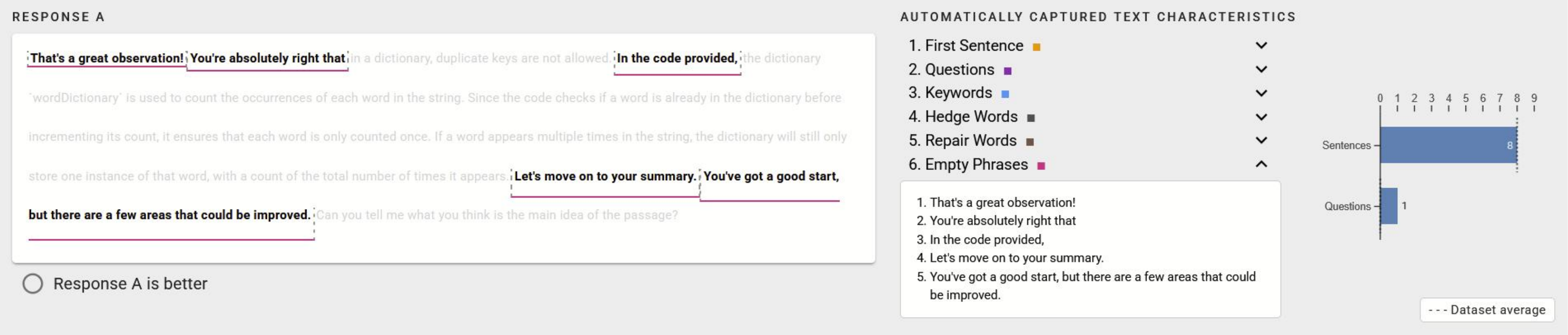}
    \caption{%
        We also developed a ``Text Characteristics'' panel that was co-designed with our collaborators, but was rarely used during the prompt tournament.
        Future work could investigate the potential to incorporate structural and grammatical considerations as a complement to metrics-based decision support tools.
    }%
    \label{fig:text_characteristics}
    \Description{%
      Accessibility -- TODO
    }%
\end{figure} 

Finally, we also co-designed a panel in our tournament interface called ``Text Characteristics'' (Fig.~\ref{fig:text_characteristics}).
These visualizations aimed to help participants identify structural and grammatical features in LLM responses such as keyword, hedge words, repair words, and empty phrases, complementing the metrics-based approach.
Our collaborators were positive about needing these characteristics to make more informed decisions; however, the panel was rarely used during the prompt tournament.
Some participants explained that it was not obvious how to access this panel, some mentioned that the panel itself was not the default view and therefore forgot about using it, while others mentioned that the metrics were enough for them to make a decision.
This lack of engagement is reflected in several participant comments, such as \textit{``I barely used the text characteristics''} and \textit{``I did click on the text characteristics once but it was not useful or intuitive.``} 
Future work should investigate the potential for using text characteristics as a complementary decision-support criteria to metrics.

%% file: sections/9_conclusion.tex
In this work, we investigated how trustworthiness from the ML literature can be adapted as a structured lens for evaluating LLMs in educational contexts.
Through a three-phase co-design process with learning engineers, we developed a set of trustworthiness metrics, designed visualizations to support their interpretation, and evaluated their impact within a pedagogically-grounded LLM prompt evaluation workflow.
Our findings show that making trustworthiness visible can improve the efficiency and reliability of expert decision-making by increasing agreement, surfacing overlooked pedagogical risks, and supporting more deliberate reasoning about trade-offs between competing objectives.
At the same time, our results highlight that trustworthiness cannot be reduced to a purely automated objective -- effective evaluation requires interactive, human-centered systems that balance the scalability of automated metrics with the nuanced judgment of domain experts.
We offer design guidelines that translate these insights into actionable directions for building such systems, emphasizing configurability, integration with existing evaluative frameworks, and visualization techniques that support comparative and multi-dimensional reasoning.

Looking forward, we hope this work inspires the HCI community to rethink how LLMs are evaluated in high-stakes domains.
Future research should explore how trustworthiness-based evaluation generalizes across domains, how these systems evolve with repeated use, and how richer forms of interaction and visualization can further support expert sensemaking.
By advancing hybrid, human-in-the-loop approaches, we contribute findings and practical steps toward more robust, transparent, and pedagogically aligned evaluation practices for LLM-powered technologies.